\documentclass{emulateapj}
\usepackage{natbib}
\usepackage{environ}
\usepackage{xcolor}

\def\apjl{{ApJ}}%
\def\apjs{{ApJS}}%
\def\aap{{A\&A}}%
\newcommand{\about}{$\sim\!\!$~}
\newcommand{\kms}{\,km\,s$^{-1}$}
\newcommand{\mstellar}{\ensuremath{M_{\mathrm{stellar}}}}
\newcommand{\omatter}{\ensuremath{\Omega_{\mathrm{M}}}}

\newcommand{\nii}{\ensuremath{\mathrm{N}\,\textsc{ii}}}
\newcommand{\sii}{\ensuremath{\mathrm{S}\,\textsc{ii}}}
\newcommand{\Siii}{\ensuremath{\mathrm{Si}\,\textsc{ii}\,\lambda6355}}
\newcommand{\vsiii}{\ensuremath{v_{\mathrm{Si}\,\textsc{ii}}}}
\newcommand{\ugriz}{\protect\hbox{$ugriz$} }
\newcommand{\deltam}{\ensuremath{\Delta m_{15}}}

\shorttitle{Hosts of high-velocity SNe~Ia}
\shortauthors{Y.-C.~Pan}

\begin{document}

\title{High-Velocity Type~Ia Supernova Has a Unique Host Environment} 

\author{
{Yen-Chen~Pan}\altaffilmark{1,2}}

\altaffiltext{1}{
Division of Science, National Astronomical Observatory of Japan, 2-21-1 Osawa, Mitaka, Tokyo 181-8588, Japan
}
\altaffiltext{2}{
EACOA Fellow
}

\begin{abstract}
Ejecta velocity of type Ia supernovae (SNe~Ia) is one powerful tool to differentiate between progenitor scenarios and explosion mechanisms. Here we revisit the relation between photospheric \Siii\ velocities (\vsiii) and host-galaxy properties with \about280 SNe~Ia. A more stringent criterion on the phase of SN spectra is adopted to classify SNe~Ia in terms of their photospheric velocities. We find significant trend that SNe~Ia with faster \Siii\ (high-\vsiii\ SNe~Ia) tend to explode in massive environments, whereas their slower counterparts can be found in both lower-mass and massive environments. This trend is further supported by the direct measurements on host gas-phase metallicities. We suggest this relation is likely caused by at least two populations of SNe~Ia. Since stars of higher metallicity (at a given mass) generally form less massive white dwarfs (WDs), our results support some theoretical models that high-\vsiii\ SNe~Ia may originate from sub-Chandrasekhar class of explosions.  Previous observations also showed some evidence that high-\vsiii\ SNe~Ia could be related to the single degenerate systems. However, we find high-\vsiii\ SNe~Ia do not come from particularly young populations. We conclude metallicity is likely the dominant factor in forming high-\vsiii\ SNe~Ia. This also implies their potential evolution with redshift and impact on the precision of SN~Ia cosmology.
\end{abstract}


\section{Introduction}
\label{sec:introduction}
\begin{figure*}
	\centering
	\begin{tabular}{c}
		\includegraphics*[scale=0.14]{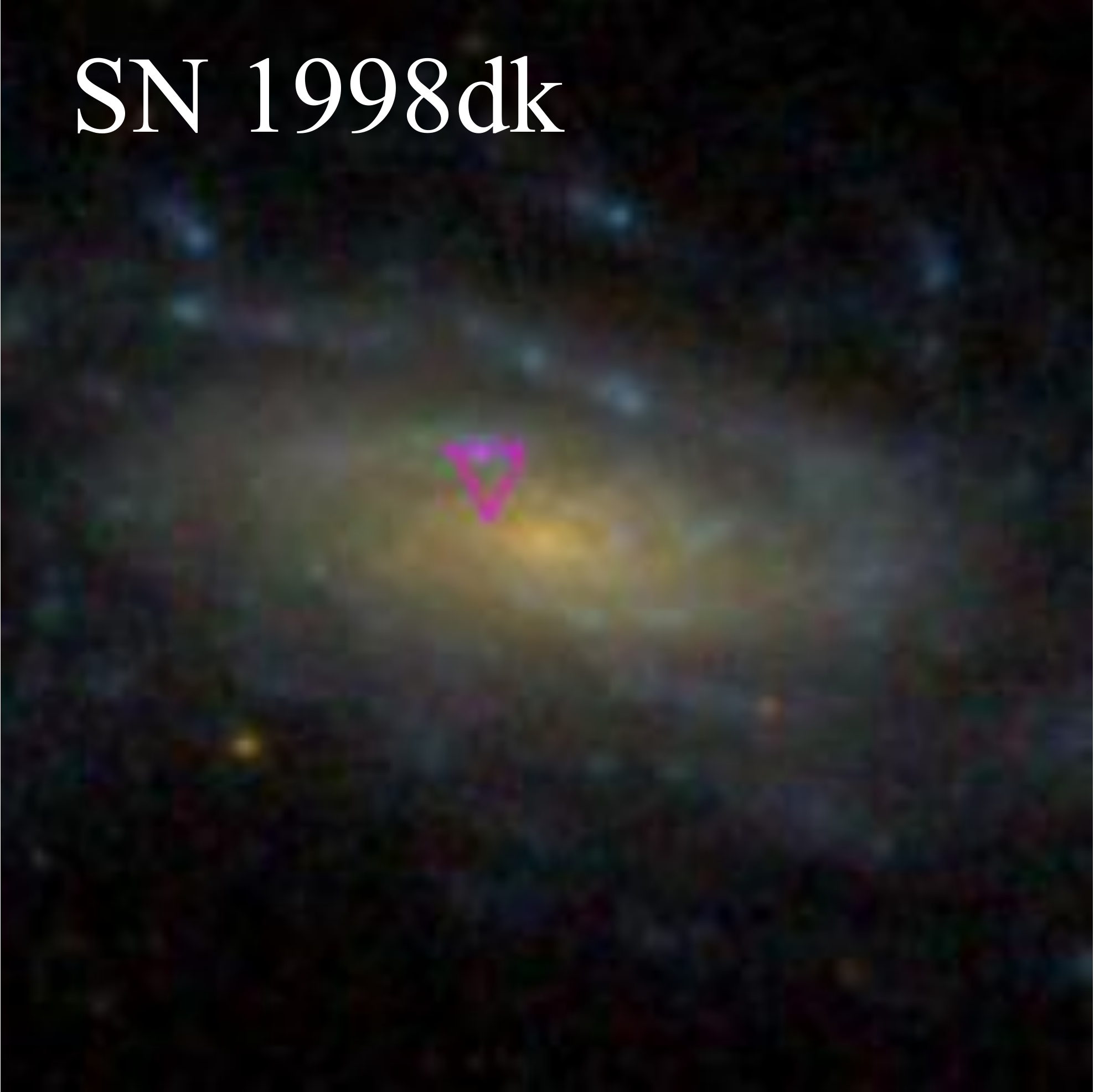}
		\includegraphics*[scale=0.14]{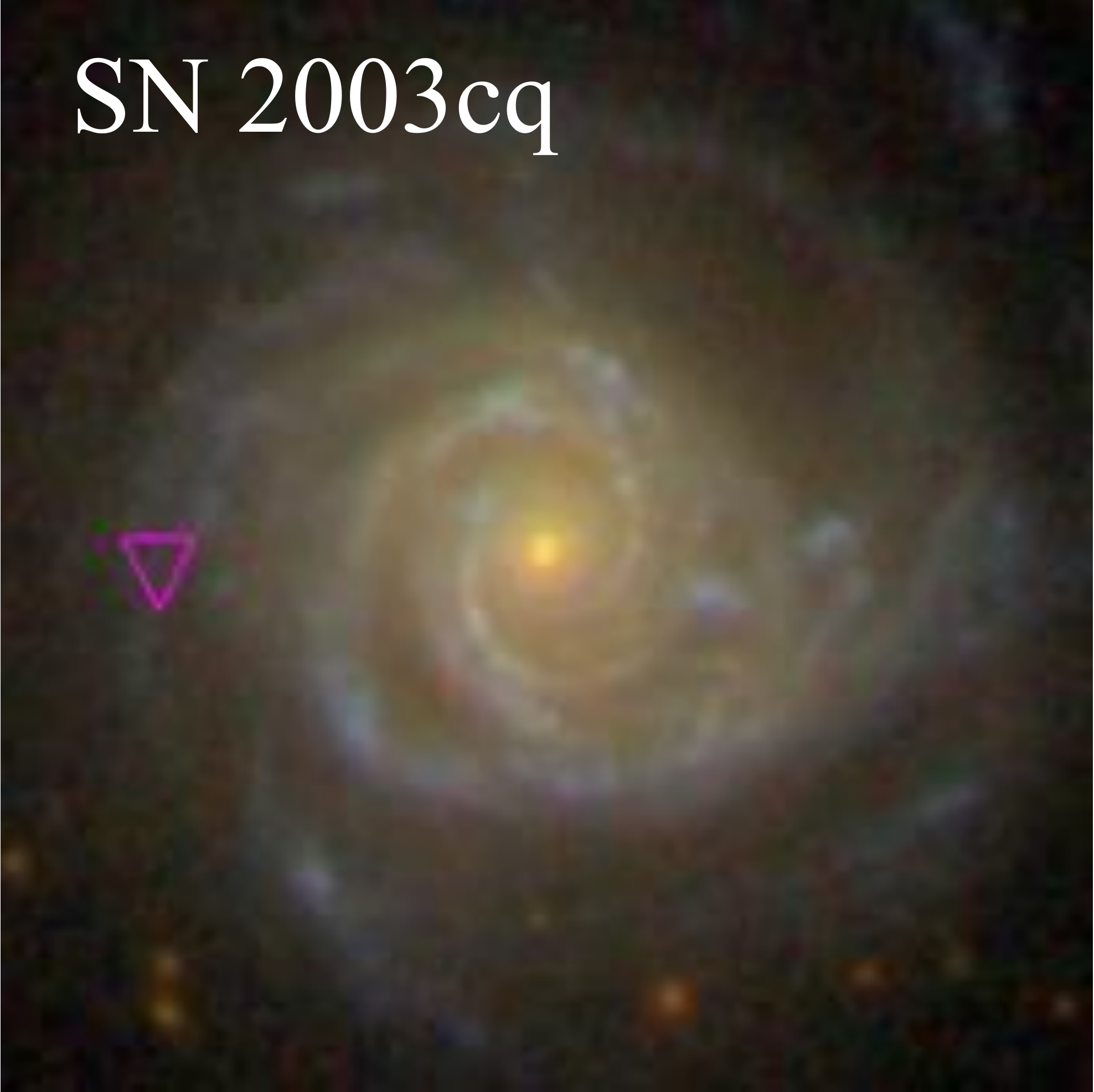}
		\includegraphics*[scale=0.14]{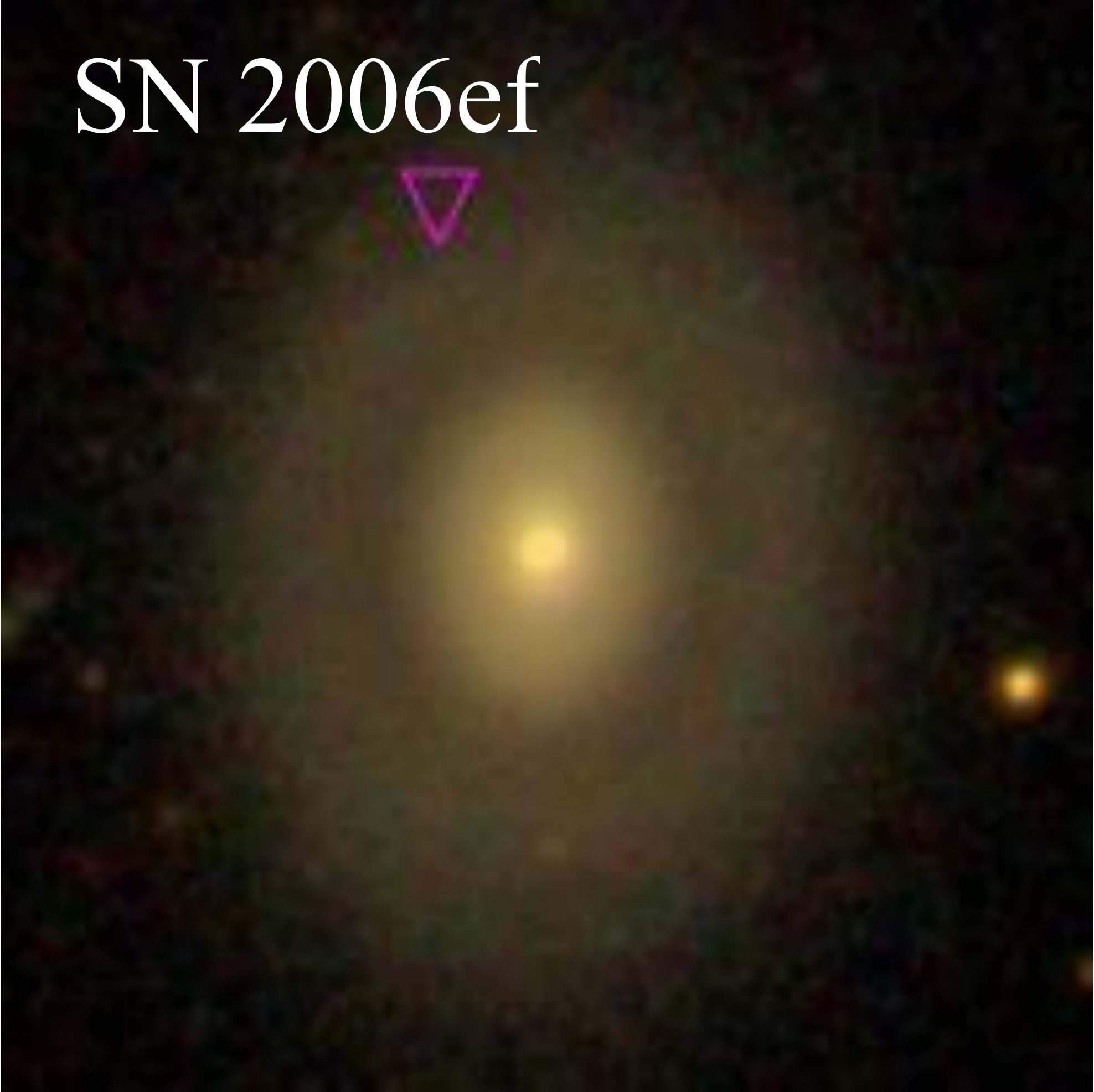}
		\includegraphics*[scale=0.14]{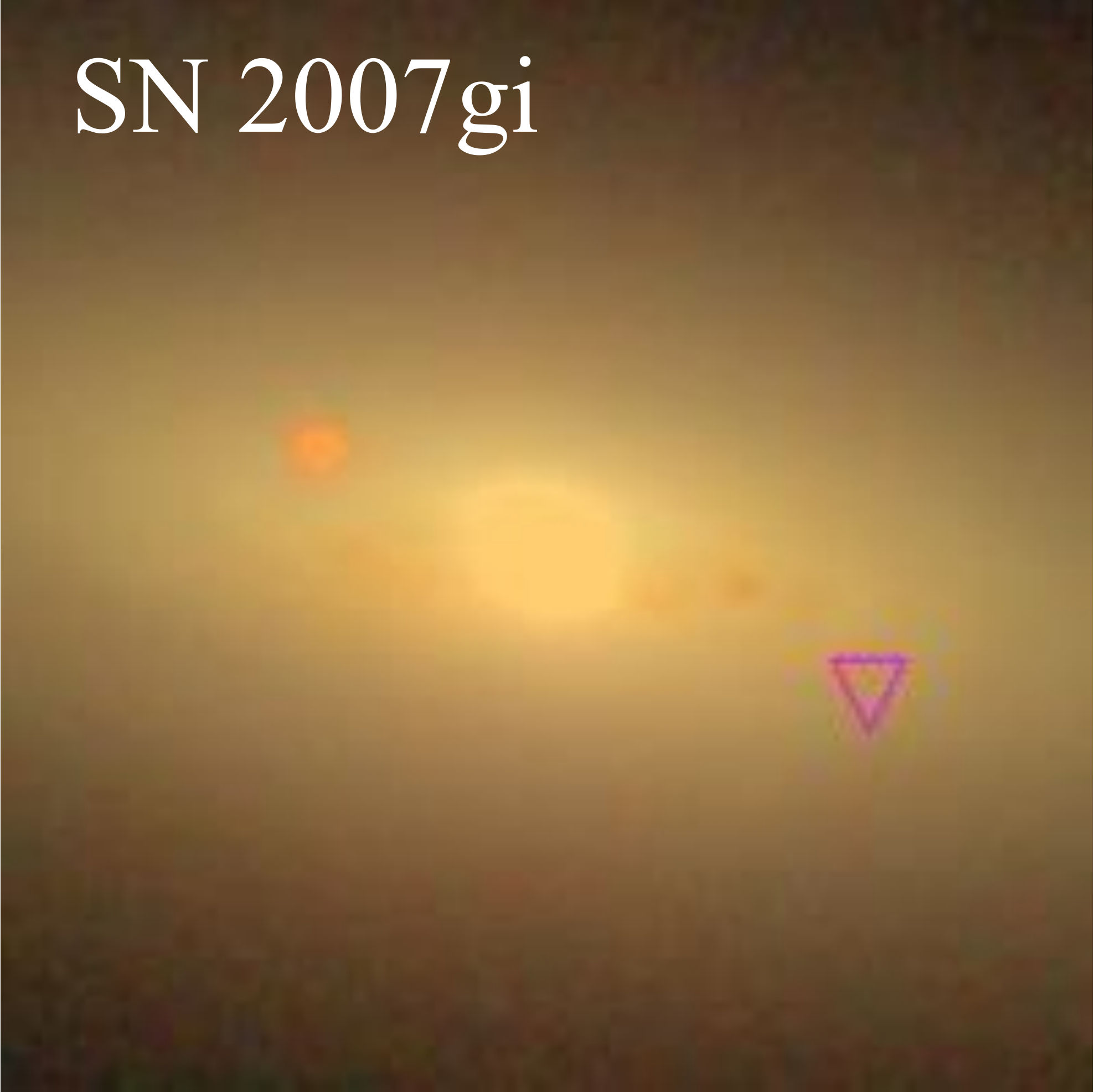}
		\includegraphics*[scale=0.14]{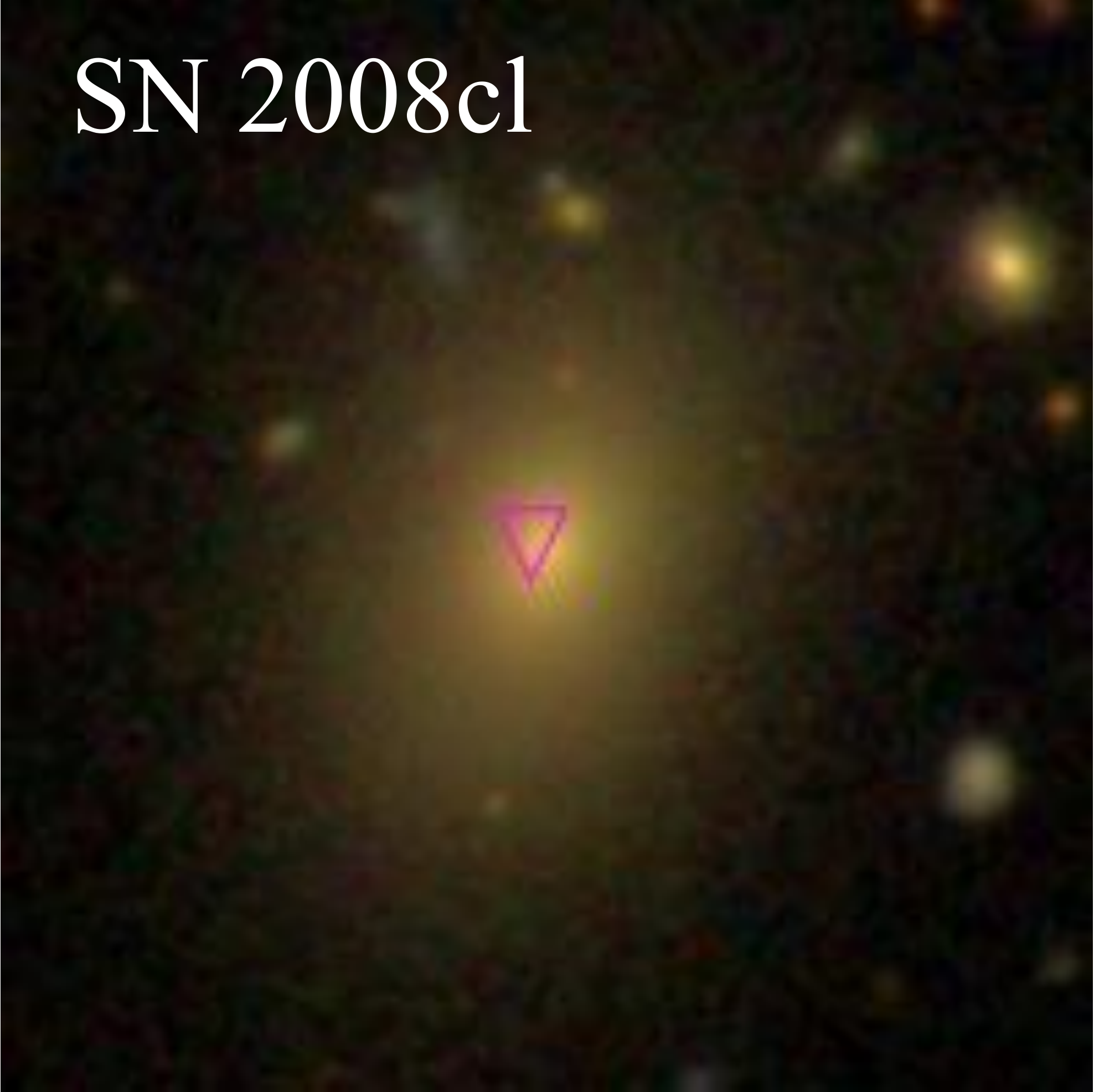}
		\includegraphics*[scale=0.14]{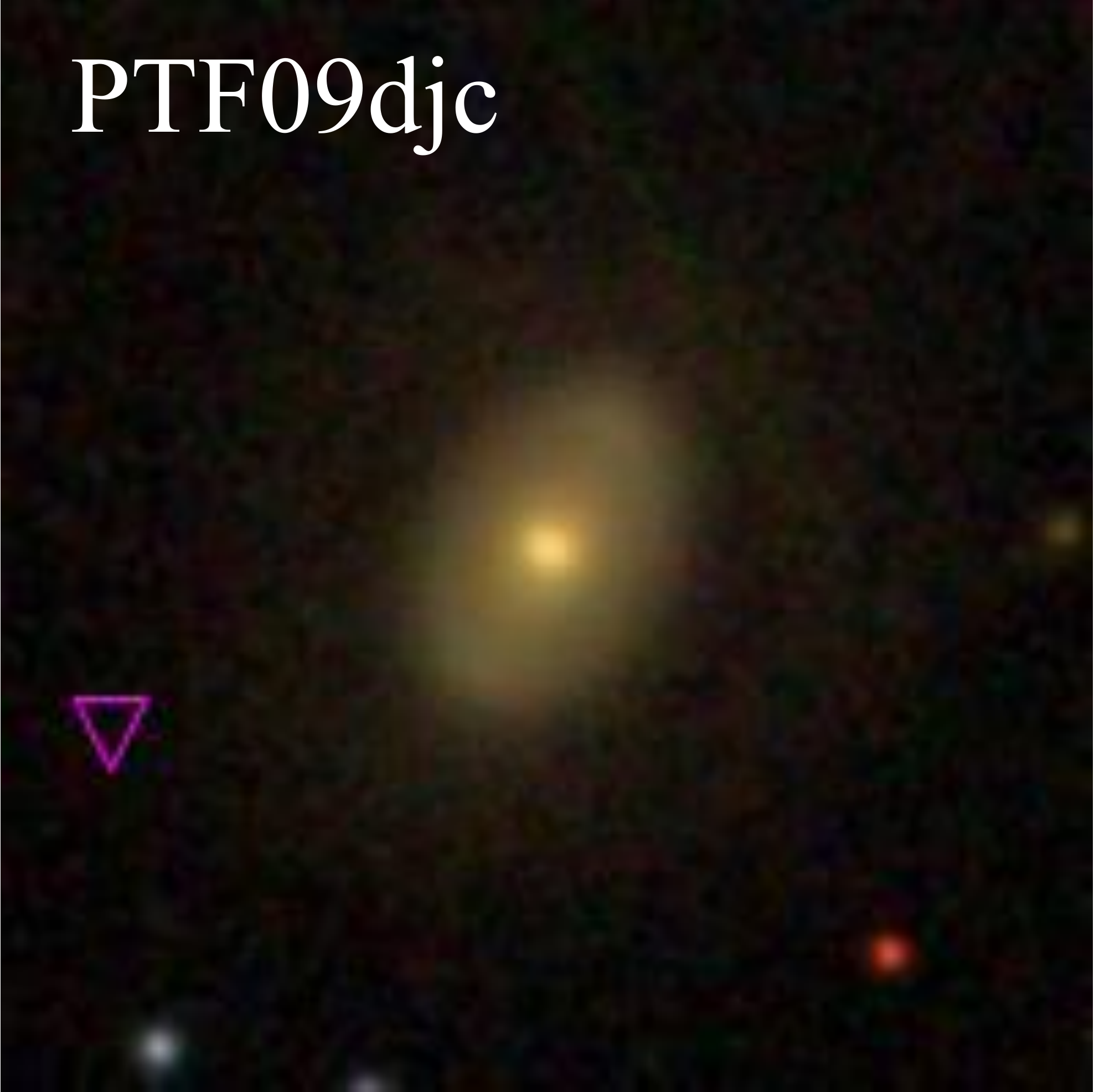}\\
		\includegraphics*[scale=0.14]{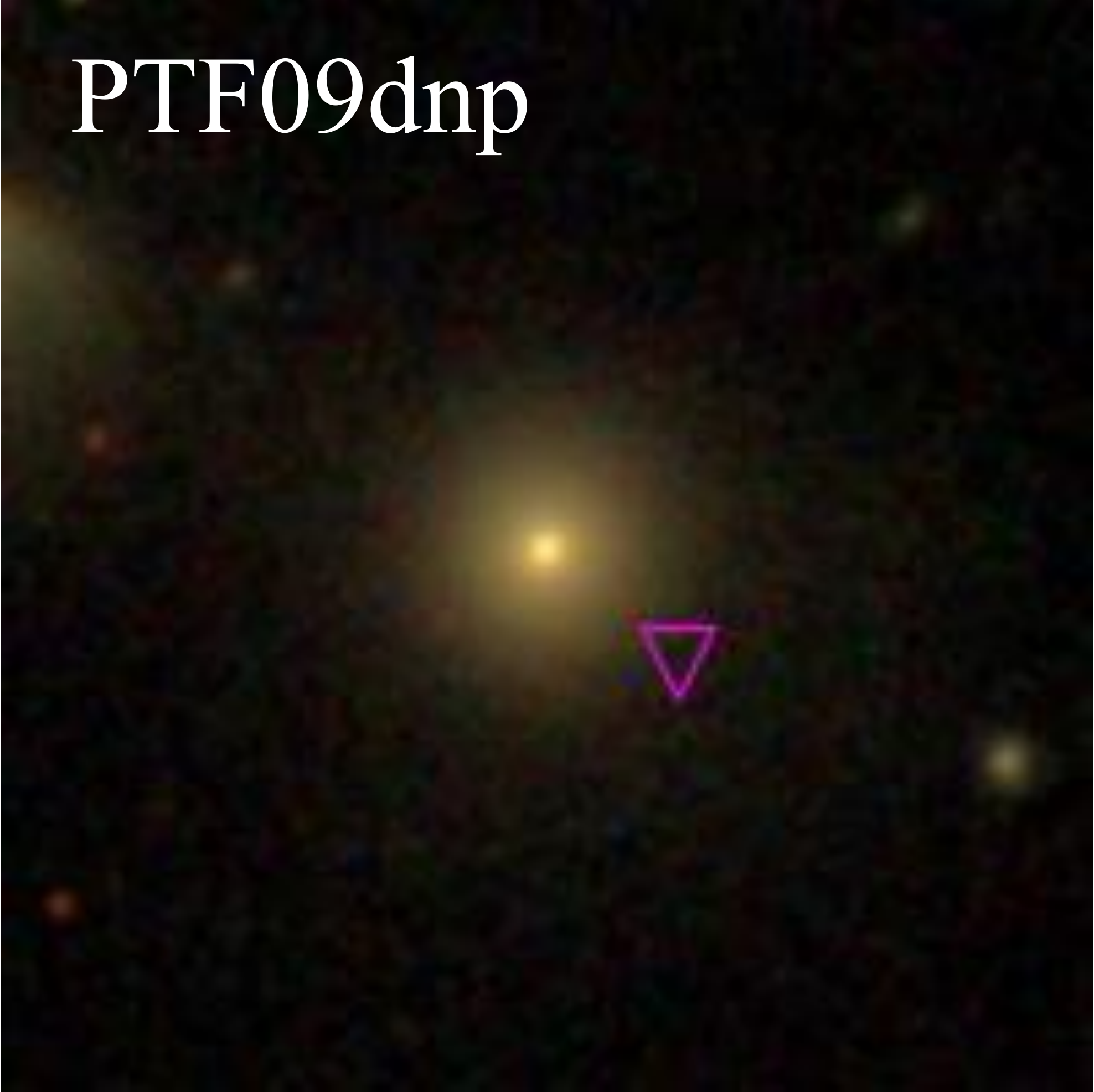}
		\includegraphics*[scale=0.14]{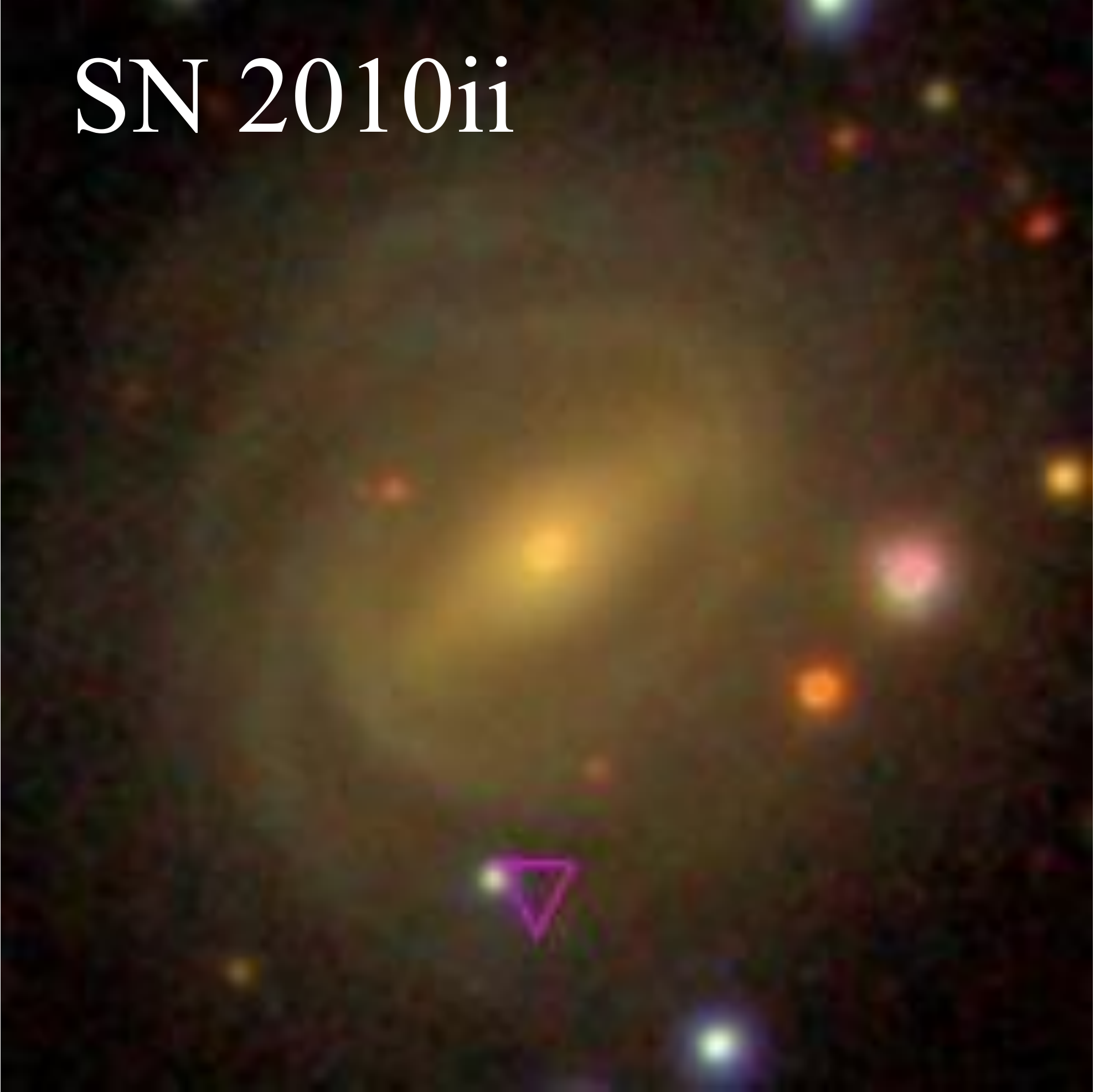}
		\includegraphics*[scale=0.14]{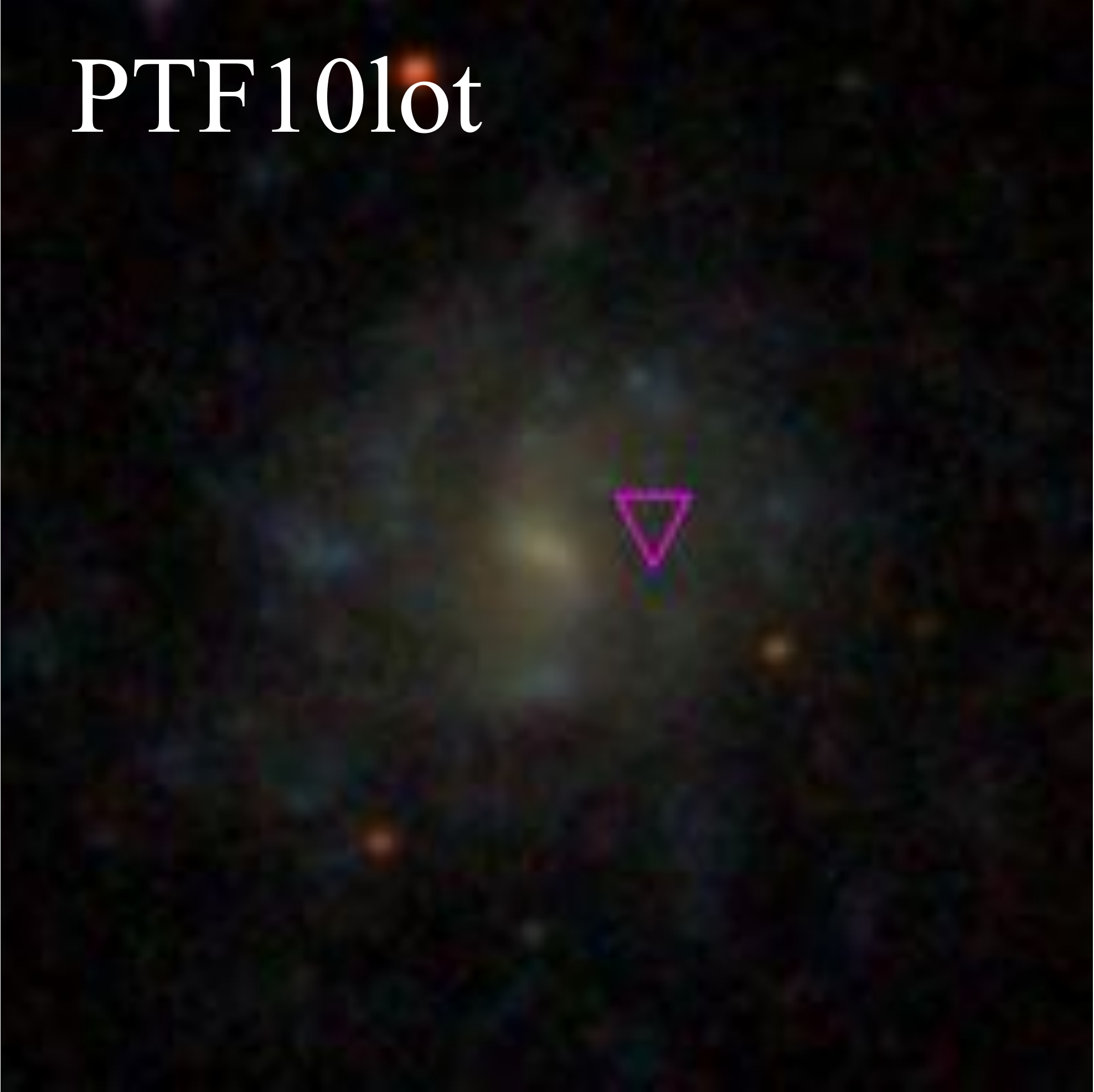}
		\includegraphics*[scale=0.14]{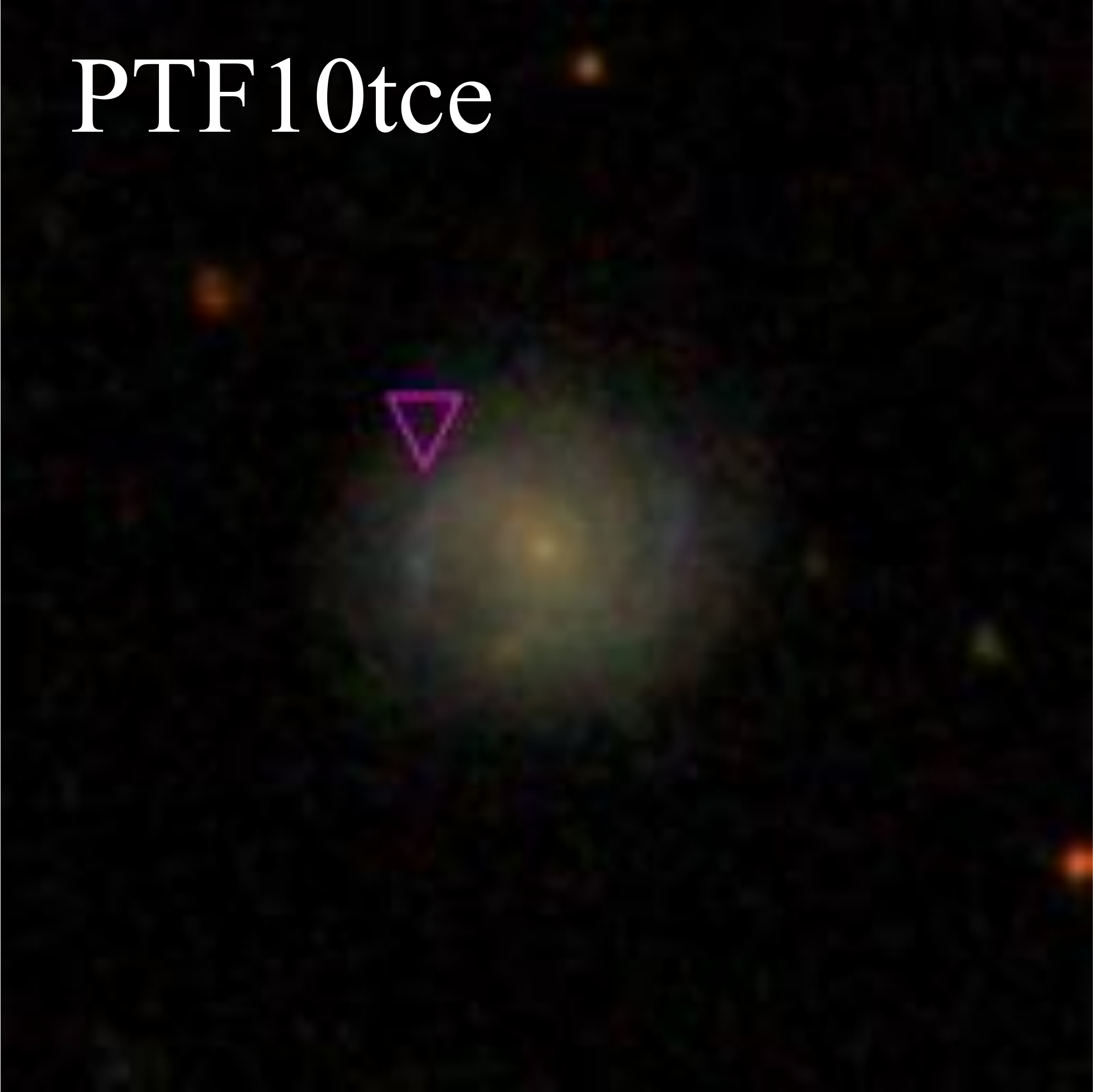}
		\includegraphics*[scale=0.14]{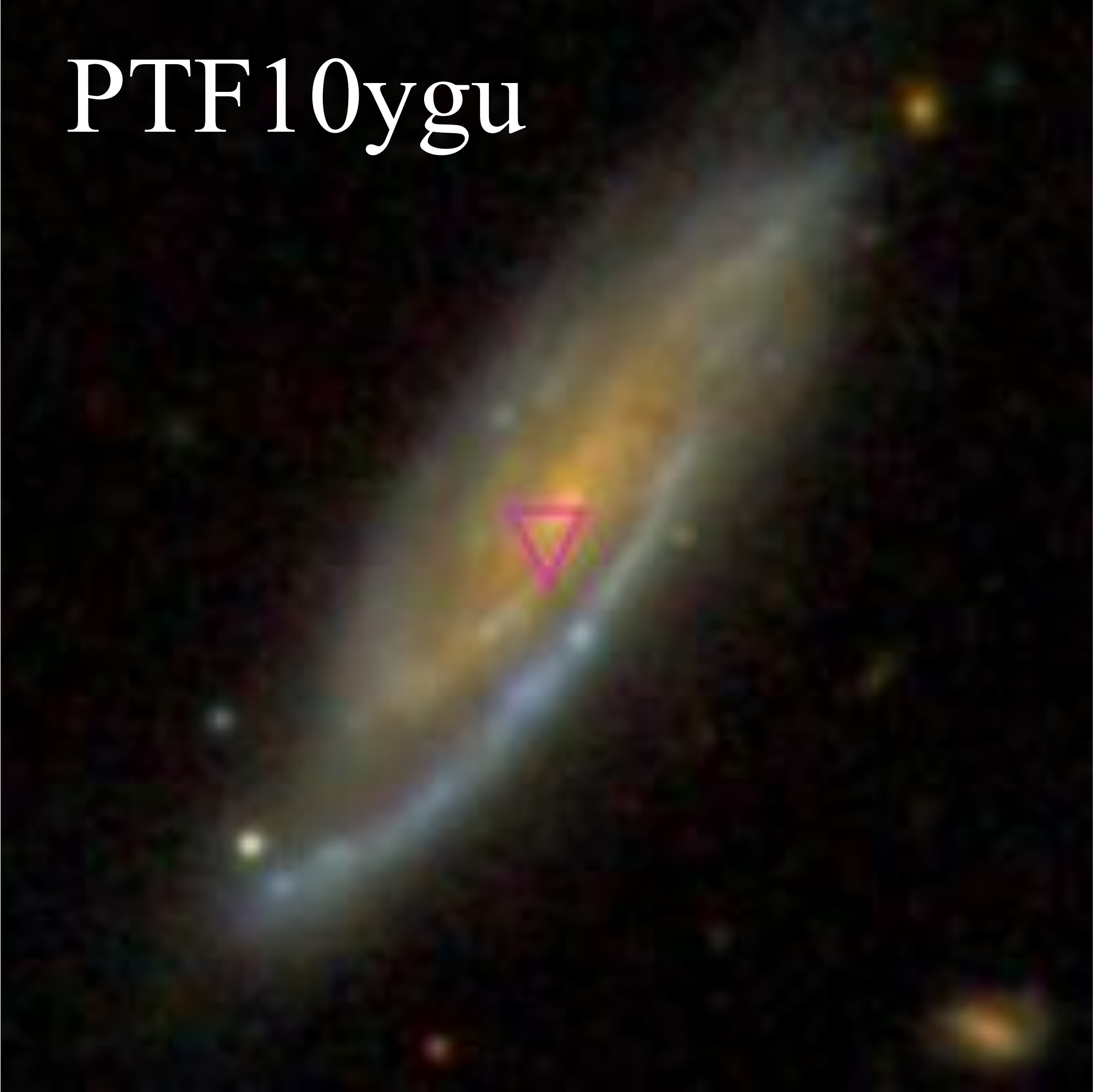}
		\includegraphics*[scale=0.14]{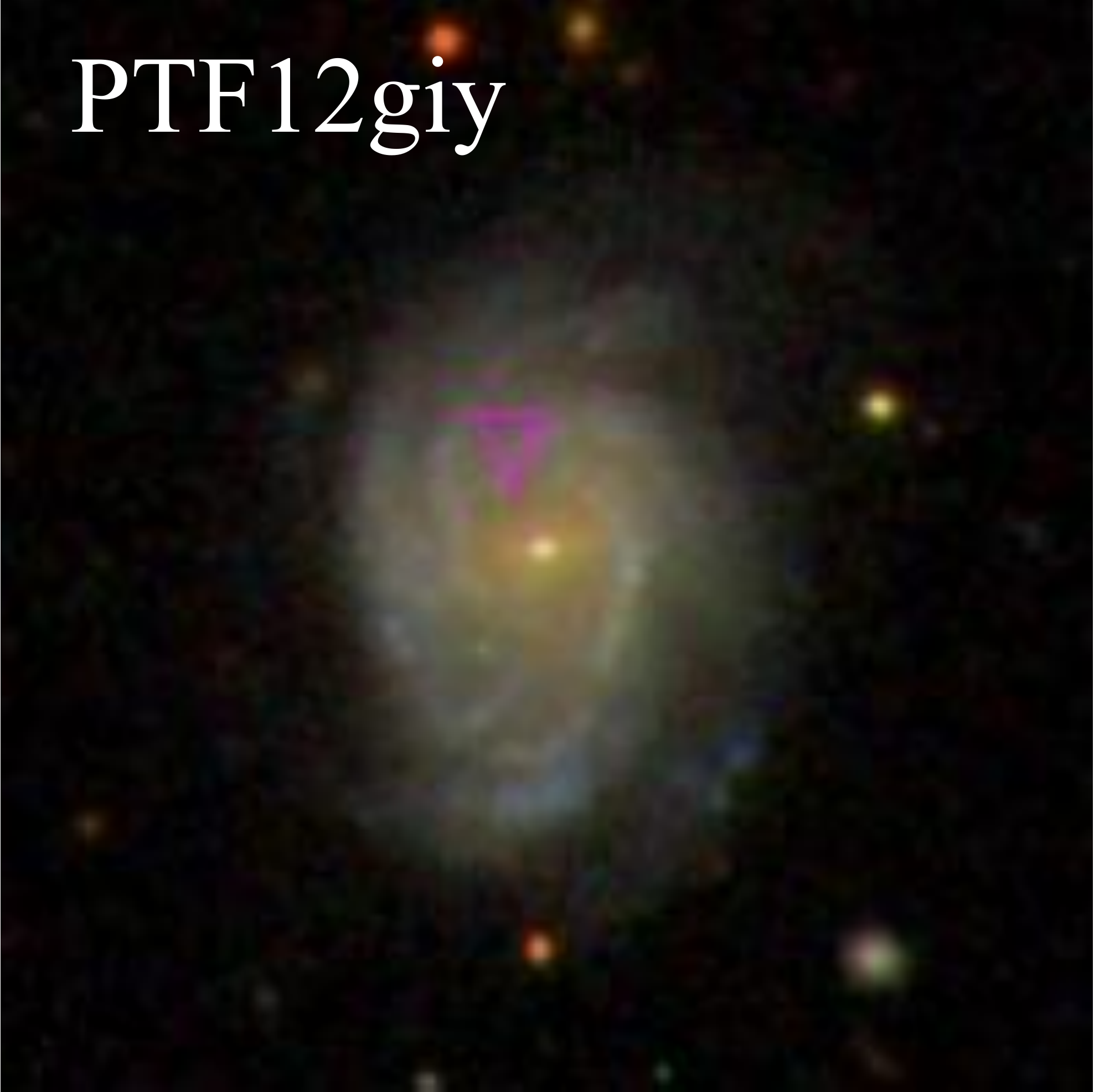}
	\end{tabular}
        \caption{Some examples of high \Siii\ velocity SN~Ia 
        (high-\vsiii\ SN~Ia; defined as $\vsiii\geq12000\,\mathrm{km\,s^{-1}}$)
        host galaxies in this work. The position of the SN is indicated by the purple triangle. 
        Images are all generated from SDSS with a size of $80\arcsec\times80\arcsec$. 
        North is up and east is left.
        }
        \label{host-cutout}
\end{figure*}

Type Ia supernovae (SNe Ia) are believed to be the result of the thermonuclear explosion of an accreting carbon-oxygen white dwarf (WD) star in a close binary system \citep{2011Natur.480..344N,2012ApJ...744L..17B}. However, the nature of the companion star that donates material is not yet clear. The various possibilities include the single degenerate \citep{1973ApJ...186.1007W} and double degenerate \citep{1984ApJS...54..335I,1984ApJ...277..355W} scenarios, as well as more variations on these themes \citep[for recent reviews, see][]{2000ARA&A..38..191H, 2013FrPhy...8..116H, 2014ARA&A..52..107M}.

The host galaxies of SNe Ia has long been a profitable route to probe the SN Ia population, with the observed properties of SNe Ia known to correlate with the physical parameters of their host galaxies. Previous studies have found significant correlations between SN Ia light curve parameters and luminosities, and the properties of their host galaxies \citep[e.g.,][]{2010ApJ...715..743K,2010ApJ...722..566L,2010MNRAS.406..782S,2011ApJ...743..172D,2013MNRAS.435.1680J,2013ApJ...770..108C,2014MNRAS.438.1391P}. SNe Ia with faster light curves are preferentially resided in massive and metal-rich galaxies than those in lower-mass and metal-poor systems. Galaxies with stronger star-formation and younger populations tend to host slower and brighter SNe Ia than passive and older galaxies. 

Previous studies also showed some evidence that the SN~Ia spectral features correlate with the host properties. By dividing SNe~Ia into two sub-groups according to their photospheric velocities, \citet{2013Sci...340..170W} found SNe Ia with high \Siii\ velocities \citep[high-\vsiii\ SNe~Ia, defined as $\vsiii \ga 12,000$\,km\,s$^{-1}$;][]{2009ApJ...699L.139W} tend to be more concentrated in the inner regions of their host galaxies, whereas the normal-velocity events (normal-\vsiii\ SNe~Ia; defined as $\vsiii < 12,000$\,km\,s$^{-1}$) span a wider range of radial distance. Given the metallicity gradients observed in both the Milky Way and many external galaxies \citep[e.g.,][]{1999PASP..111..919H}, they suggested the metallicity could be important in driving this relation. \citet[][hereafter P15]{2015MNRAS.446..354P} further supported this idea by finding some evidence that high-\vsiii\ SNe~Ia tend to reside in more massive galaxies than the normal-\vsiii\ counterparts, although their results are not statistically significant due to the small sample size. These host studies together suggested there could be at least two distinct populations of SNe~Ia in terms of their ejecta velocities.

High-\vsiii\ SN~Ia has long been suspected to have a different origin from normal-\vsiii\ SN~Ia. This was firstly proposed by \citet{2009ApJ...699L.139W}, where they found high-\vsiii\ SNe~Ia tend to be redder and prefer a lower extinction ratio ($R_{V}$) than normal-\vsiii\ SNe~Ia. \citet{2011ApJ...729...55F} argued the high-\vsiii\ SNe~Ia actually do not have a different reddening law but are intrinsically redder than normal-\vsiii\ SNe~Ia. Recently, \citet{2019ApJ...882..120W} found some evidence that these high-\vsiii\ SNe~Ia tend to show blue excess in their late-time light-curves and variable Na\,\textsc{i} absorption lines in their spectra. These observations were attributed as the circumstellar dust surrounding the SNe. They claimed the high-\vsiii\ SNe~Ia are likely associated with the single degenerate systems.

Theoretical study also suggested the high-\vsiii\ SNe~Ia may originate from unique explosions. Using 1D WD explosion models, \citet{2019ApJ...873...84P} showed the sub-Chandrasekhar explosions could produce SNe~Ia with a wide range of \vsiii. Their results further indicated the high-\vsiii\ SNe~Ia could be primarily produced by sub-Chandrasekhar type of explosions. The red color found for high-\vsiii\ SNe~Ia can also be explained by the line blanketing effect due the ashes of helium shell in their models.

In this paper, we revisit the relation between SN~Ia \vsiii\ and host-galaxy properties with a parent sample of \about280 SNe~Ia ($\gtrsim$2 times larger than that in P15), with the purpose to differentiate the progenitor properties and explosion mechanisms between high-\vsiii\ and normal-\vsiii\ SNe~Ia. A plan of the paper follows. In Section~\ref{sec:data} we introduce our SN Ia spectral sample and the determination of host parameters. We show the results in Section~\ref{sec:analysis}. The discussion and conclusions are presented in Section~\ref{sec:discussion} and Section~\ref{sec:conclusions}, respectively. Throughout this paper, we assume $\mathrm{H_0}=70$\,km\,s$^{-1}$\,Mpc$^{-1}$ and a flat universe with $\omatter=0.3$.

\section{Data and method}
\label{sec:data}
\begin{figure*}
	\centering
	\begin{tabular}{c}
		\includegraphics*[scale=0.62]{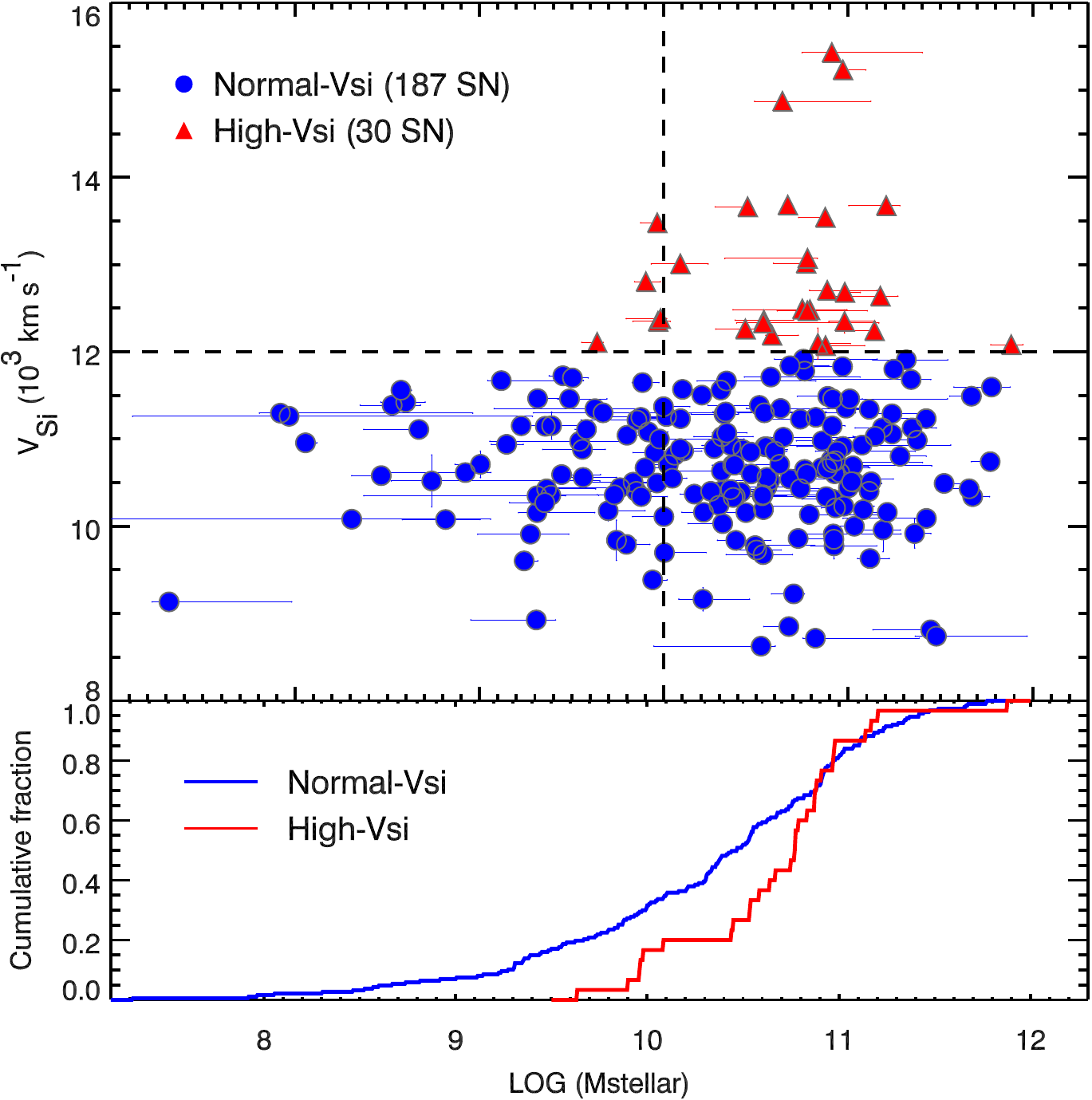}
		\hspace{0.25cm}
		\includegraphics*[scale=0.62]{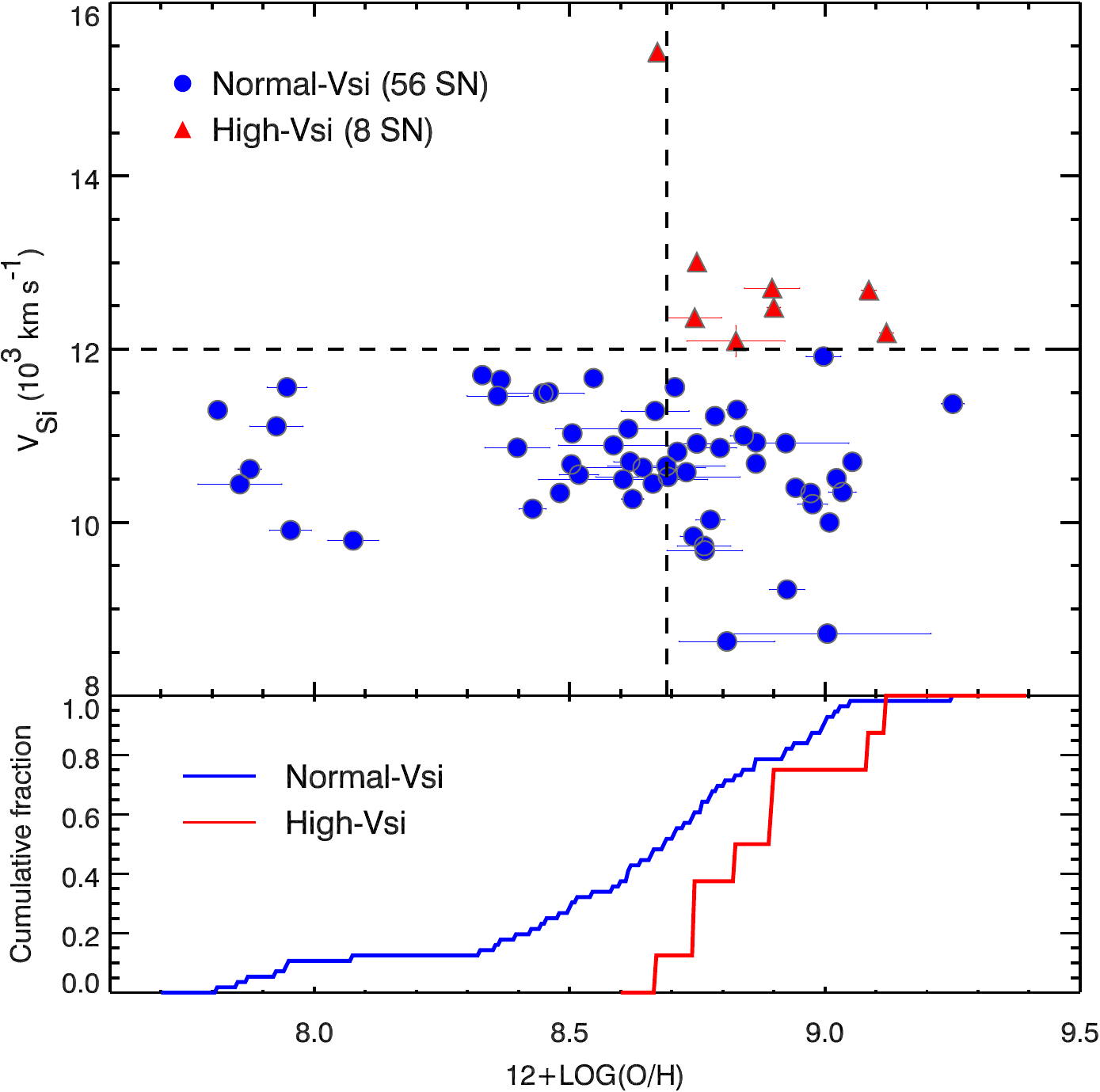}
	\end{tabular}
        \caption{{\it Left}: The \Siii\ velocities (\vsiii) as a function
                  of host-galaxy stellar mass (\mstellar).  The high-\vsiii\ SNe Ia
                  are shown as red triangles, and the normal-\vsiii\ SNe Ia are
                  shown as blue circles. The vertical and horizontal dashed
                  lines represent the criterion used to split the
                  sample in velocity and \mstellar\ space,
                  respectively. The bottom histograms show
                  the cumulative fractions of \mstellar\ for high-\vsiii\ and
                  normal-\vsiii\ SNe Ia. {\it Right}: The same as left panel,
                  but with host-galaxy gas-phase metallicity instead.}
        \label{mass-vsi6150}
\end{figure*}
\subsection{SN data}
\label{sec:sn-data}
In this work, we extend our analysis with the full spectroscopic sample studied in \citet[][hereafter M14]{2014MNRAS.444.3258M}. This is the parent sample of what was studied by P15, containing 264 spectroscopically normal SNe~Ia with \Siii\ measurements near peak (e.g., within 5 days from the peak luminosity). They were all discovered by the Palomar Transient Factory (PTF). In addition to the PTF sample, we further include the SNe studied in \citet[][hereafter S15]{2015MNRAS.451.1973S} to increase our sample size. This added another \about150 SNe~Ia with near-peak \Siii\ measurements from the Berkeley SN~Ia Program (BSNIP) after removing the duplicate objects from our PTF sample. This gives a sample of \about400 SNe~Ia at $z<0.2$. The description of the spectroscopic observation and data reduction can be found in detail in M14 and S15.

Since M14 and S15 used very similar techniques in measuring the spectral features, we do not perform new measurements but simply adopting their results in our analysis. A complete description of the line measurement can be found in M14. Briefly speaking, the SN spectrum is firstly corrected into the rest frame, define (by eye inspection) continuum regions on either side of the feature, and fit a straight line pseudo-continuum across the absorption feature. The feature is then normalised by dividing it by the pseudo-continuum. A Gaussian fit is performed to the normalised \Siii\ line in velocity space. The resulting fit then gives the velocity and pseudo-equivalent widths (pEW) of the feature. 

\subsection{Host-galaxy properties}
\label{sec:host}
The main purpose of this work is to investigate the relation between \vsiii\ of SNe~Ia and their host properties. The host stellar mass (\mstellar) is derived by fitting the photometry of the host galaxy with the photometric redshift code \textsc{z-peg} \citep{2002A&A...386..446L}. The host photometry is provided from SDSS \ugriz catalog \citep{2018ApJS..235...42A}. The SDSS model magnitudes are used here. \textsc{z-peg} fits the observed galaxy colours with galaxy spectral energy distribution (SED) templates corresponding to 9 spectral types (SB, Im, Sd, Sc, Sbc, Sb, Sa, S0, and E). Here we assume a \citet{1955ApJ...121..161S} initial-mass function (IMF). The photometry is corrected for foreground Milky Way reddening with $R_{V} = 3.1$ and a \citet*[][CCM]{1989ApJ...345..245C} reddening law. Fig.~\ref{host-cutout} shows some SDSS color images of high-\vsiii\ SN~Ia host galaxies studied in this work.

We also measure the host gas-phase metallicity and star-formation rate (SFR). This is done by obtaining the optical spectra of the host galaxies, primarily with the SDSS spectrograph on the Sloan Foundation 2.5-m telescope and Gemini Multi-Object Spectrographs (GMOS) on the Gemini Observatory. We fit the emission lines and stellar continuum of the host spectrum using the Interactive Data Language (\textsc{IDL}) codes \textsc{ppxf} \citep{2004PASP..116..138C} and \textsc{gandalf} \citep{2006MNRAS.366.1151S}. A complete description of this process can be found in \citet{2014MNRAS.438.1391P}. Briefly, \textsc{ppxf} fits the line-of-sight velocity distribution (LOSVD) of the stars in the galaxy in pixel space using a series of stellar templates. Before fitting the stellar continuum, the wavelengths of potential emission lines are masked to remove any possible contamination. The stellar templates are based on the \textsc{miles} empirical stellar library \citep{2006MNRAS.371..703S, 2010MNRAS.404.1639V}.  A total of 288 templates is selected with $[M/H]=-1.71$ to $+0.22$ in 6 bins and ages ranging from $0.063$ to $14.12$\,Gyr in 48 bins.

We correct all spectra for foreground Galactic reddening using the calibrations of \citet{2011ApJ...737..103S}. The host-galaxy extinction is corrected with the two-component reddening model in \textsc{gandalf}. The first component assumes a diffusive dust throughout the whole galaxy that affects the entire spectrum. It is determined by comparing the observed spectra to the un-reddened spectral templates. The second component measures the local dust around the nebular regions and affects only the emission lines. It is constrained only if the Balmer decrement (the H$\alpha$ to H$\beta$ line ratio) can be measured.

After the emission-line measurements from \textsc{ppxf} and \textsc{gandalf}, we determine the SFR by adopting the conversion of \citet{1998ARA&A..36..189K}, which used evolutionary synthesis models to relate the luminosity of the H$\alpha$ line to the SFR. We calculate the host gas-phase metallicity based on the diagnostics from \citet{2016Ap&SS.361...61D}. \citet{2016Ap&SS.361...61D} used the ratios of [\nii]\,$\lambda6584$ to [\sii]\,$\lambda\lambda6717,6731$ and [\nii]\,$\lambda6584$ to H$\alpha$ to calibrate the gas-phase metallicity. This has the advantage of requiring a narrow wavelength range and therefore is less affected by the reddening correction. We further use BPT diagrams \citep*{1981PASP...93....5B} to check for potential contamination from active galactic nuclei (AGNs) in our host galaxies. The criteria proposed by \citet{2001ApJ...556..121K} are adopted to distinguish between normal and AGN host galaxies. The potential AGNs are excluded from our emission line analyses. A summary of our measurements can be found in Table~\ref{sample1}.

\section{Analysis}
\label{sec:analysis}
We firstly investigate the relation between \vsiii\ and host \mstellar. Ideally, one should distinguish between high-\vsiii\ and normal-\vsiii\ SNe~Ia with the \vsiii\ measured exactly at peak luminosity. In practice, previous studies generally used the spectra within a few days from the peak given that the spectral evolution is relatively mild at those epochs. M14 measured \vsiii\ using all the spectra observed within 5 days from the peak luminosity and did not apply phase corrections to their line measurements. Given a larger sample size, we adopt a more stringent phase criterion in this work; we use all the spectra observed within only 3 days from the peak luminosity. That way the \vsiii\ of different SNe are compared at closer phases and reduces the uncertainties of phase evolution. This gives a final parent sample of 281 SNe (41 of them are high-\vsiii\ SNe~Ia), which is still more than 2 times larger than that studied in P15. Releasing this criterion does not change our conclusion but only making our results less significant.

The result is shown in the left panel of Fig.~\ref{mass-vsi6150}. We find clear evidence that most of the high-\vsiii\ SNe~Ia reside in massive galaxies ($\log(\mstellar)>10\,M_{\odot}$), whereas the normal-\vsiii\ SNe~Ia can be found in both lower-mass and massive galaxies. This is consistent with the finding in P15. Here we confirm this trend with a much larger sample and our results are statistically more significant. Both Kolmogorov-Smirnov (K-S) and Anderson-Darling (A-D) tests give a $p$-value of $\lesssim$0.02 that the \mstellar\ distributions of high-\vsiii\ and normal-\vsiii\ SNe~Ia are drawn from the same underlying population. This value is \about10 times smaller than that studied in P15. It is also evident that this relation is neither linear nor monotonic. In fact, it implies the existence of multiple populations of SNe~Ia (see Section~\ref{sec:population} for a discussion).

We also note the \vsiii\ of \about12\,000\,\kms\ is a fairly good criterion to distinguish between high-\vsiii\ and normal-\vsiii\ SNe~Ia. For SNe with \vsiii\ $>12\,000$\,\kms, \about84 percent of their host galaxies have $\log(\mstellar)>10\,M_{\odot}$. The ratio goes up to only \about90 percent if the criterion is raised to 13\,000\,\kms, but drops significantly to \about68 percent if the criterion is lowered to 11\,000\,\kms. All of the SNe with \vsiii\ $>12\,000$\,\kms\ have $\log(\mstellar)>9.6\,M_{\odot}$.

Next we investigate the relation with host gas-phase metallicity. The result is shown in the right panel of Fig.~\ref{mass-vsi6150}. Given the tight relation between \mstellar\ and metallicity \citep[e.g.,][]{2004ApJ...613..898T}, it is reasonable to suspect that the metallicity is the underlying source to drive the relation we see with \mstellar. In general, we find the relation between \vsiii\ and host metallicity is consistent with what we have found for host \mstellar. The host galaxies of high-\vsiii\ SNe~Ia tend to be metal-rich, having metallicities mostly above solar value \citep[8.69;][]{2001ApJ...556L..63A}. The K-S and A-D tests give a $p$-value of 0.05 and 0.04, respectively, that the metallicity distributions of high-\vsiii\ and normal-\vsiii\ SNe~Ia are drawn from the same underlying population. This is less significant than the trend with host \mstellar. However, our sample with host metallicity measurements is only one third the size of that with host \mstellar. The progenitor metallicity is also expected to differ from nuclear metallicity measurements performed in this work. A larger sample with direct metallicity measurements near the SN location is critical to constrain the metallicity effect in the future.

\section{Discussion}
\label{sec:discussion}
\subsection{Silicon velocity and metallicity}
\label{sec:population}
\citet{2000ApJ...530..966L} showed that the observed \vsiii\ could vary with the C+O layer metallicity in SN~Ia. The blue-shifted velocity of the \Siii\ feature increase with C+O layer metallicity due to the increasing opacity in the C+O layer moving the features blueward and causing larger line velocities. P15 determined a linear relation between the \vsiii\ and C+O metallicity of SN progenitor using the models of \citet{2000ApJ...530..966L} and showed the \vsiii\ increase with metallicities with a slope of 435\,\kms\,$\rm dex^{-1}$. They claimed the observed relation between host \mstellar\ and \vsiii\ is in qualitative agreement with that of \citet{2000ApJ...530..966L} models. However, fitting a linear relation between the observed \vsiii\ and host gas-phase metallicity with our sample (i.e., right panel of Fig.~\ref{mass-vsi6150}) gives a slope of $87\pm427$\,\kms\,$\rm dex^{-1}$, which is consistent with no trend.

In fact, it is now evident that both high-\vsiii\ and normal-\vsiii\ SNe~Ia can be found in metal-rich host environments. If the high photospheric velocity is mainly caused by the increasing opacity (due to higher progenitor metallicity) in SN, we would expect a monotonic relation between \vsiii\ and host metallicity, instead of a L-shaped distribution shown in both panels of Fig.~\ref{mass-vsi6150}. Thus, it is not precise to say that the high-\vsiii\ SNe~Ia tend to reside in {\it more} metal-rich environments than that of normal-\vsiii\ SNe~Ia. While the opacity could still have some effect here, we argue that there are likely at least two populations of SNe~Ia responsible for the observed trend. The high-\vsiii\ SNe~Ia could be part of a unique population which is sensitive to the progenitor metallicity and can only be formed in metal-rich environments  (see Section~\ref{sec:explosion} for a discussion).

\subsection{Implications on progenitor systems and explosion mechanisms}
\label{sec:explosion}
\begin{figure}
	\centering
	\begin{tabular}{c}
		\includegraphics*[scale=0.5]{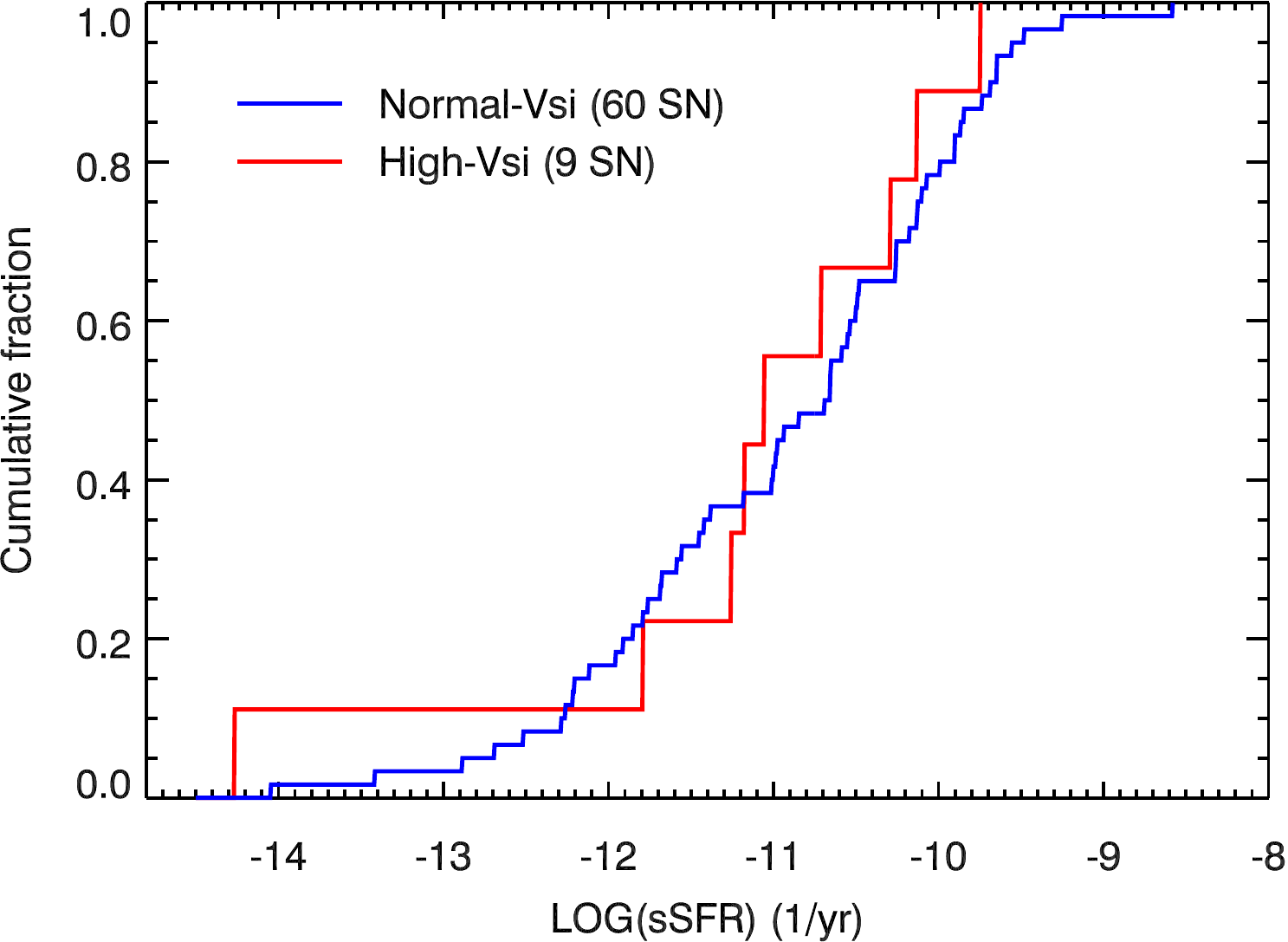}
	\end{tabular}
        \caption{The cumulative fractions of specific star-formation rate (sSFR) 
        for high-\vsiii\ (red) and normal-\vsiii\ (blue) SNe~Ia.}
        \label{ssfr-cdf}
\end{figure}

\citet{2013Sci...340..170W} found high-\vsiii\ SNe~Ia are more concentrated in the inner and brighter regions of their host galaxies. They suggested the high-\vsiii\ SNe~Ia likely originate from younger and more metal-rich progenitors than those of normal-\vsiii\ SNe~Ia. Recently, \citet{2019ApJ...882..120W} showed some evidence that these high-\vsiii\ SNe~Ia tend to present blue excess in their late-time light-curves and variable Na\,\textsc{i} absorption lines in the spectra. They attributed these observations to the circumstellar dust surrounding the SNe and concluded that the high-\vsiii\ and normal-\vsiii\ SNe~Ia are likely from the single degenerate and double degenerate systems, respectively.

Our results support that the high-\vsiii\ SN~Ia has a strong preference to occur in metal-rich environment. However, we argue that they may not come from particularly young populations. Fig.~\ref{ssfr-cdf} shows the cumulative fractions of specific star-formation rate (sSFR) for high-\vsiii\ and normal-\vsiii\ SNe Ia. Here the sSFR is defined as the SFR per unit \mstellar. Theoretically, the sSFR is a more appropriate indicator to measure the relative star-formation activity of a galaxy as it measures the star-formation relative to the underlying galaxy stellar mass \citep{1997ApJ...489..559G}. There is also a strong correlation between sSFR and age of the galaxies, in a sense that higher-sSFR galaxies tend to have younger stellar populations than lower-sSFR galaxies \citep[e.g.,][]{2004MNRAS.351.1151B}. 

We find the difference between high-\vsiii\ and normal-\vsiii\ SNe Ia is not statistically significant in terms of their host sSFR. The K-S test gives a $p$-value of 0.93 that the sSFR distributions of high-\vsiii\ and normal-\vsiii\ SNe~Ia are drawn from the same underlying population. We determine a mean $\rm\log(sSFR)$ of $-10.93\pm1.11$\,yr$^{-1}$ and $-11.15\pm1.16$\,yr$^{-1}$ for normal-\vsiii\ and high-\vsiii\ SNe~Ia, respectively. Thus, the host galaxies of high-\vsiii\ SNe~Ia do not tend to be younger than their normal-\vsiii\ counterparts. This is consistent with the results in P15, where they showed the youngest populations are likely related to those SNe~Ia with dispatched high-velocity features (HVFs), not those with high photospheric velocities. They also found there is a significant number (more than 30\,percent) of high-\vsiii\ SNe~Ia in early-type galaxies. Our results imply the metallicity is probably the only important (or dominant) factor in forming high-\vsiii\ SNe~Ia. 

Theoretical studies also suggested the high-\vsiii\ SNe~Ia may have unique explosion mechanisms. Using 1D WD explosion models, \citet{2019ApJ...873...84P} showed the sub-Chandrasekhar class of explosions can produce SNe~Ia of a wide range of luminosities and photospheric velocities. In particular, their results indicated the high-\vsiii\ SNe~Ia could be primarily produced by sub-Chandrasekhar explosions, whereas normal-\vsiii\ SNe~Ia can be produced by both sub-Chandrasekhar and near-Chandrasekhar explosions. The significant line blanketing due to the ashes of helium shell in their models also explained the intrinsically red color of high-\vsiii\ SNe~Ia \citep[e.g.,][]{2011ApJ...729...55F}. However, it is not yet clear the contribution of progenitor metallicity on such models.  

Progenitor metallicity is believed to have significant impact on explosions of SN~Ia. For example, it may affect the accretion onto the WD by changing the opacity in the wind for some single degenerate scenarios \citep[e.g.,][]{1998ApJ...503L.155K}. The mass of the WD is also expected to vary with metallicity. At a given mass, stars of higher metallicity generally produce less massive WDs \citep[e.g.,][]{1999ApJ...513..861U}. This implies they may be more difficult to reach the Chandrasekhar limit for explosions. Under the circumstances, the sub-Chandrasekhar class of explosions are probably more efficient and may account for some SNe~Ia having higher progenitor metallicities. The preference of metal-rich environments can be used as a strong constraint to discriminate between models for high-\vsiii\ SNe~Ia in the future.

\subsection{Implications on cosmology}
\label{sec:cosmology}
\begin{figure}
	\centering
	\begin{tabular}{c}
		\includegraphics*[scale=0.5]{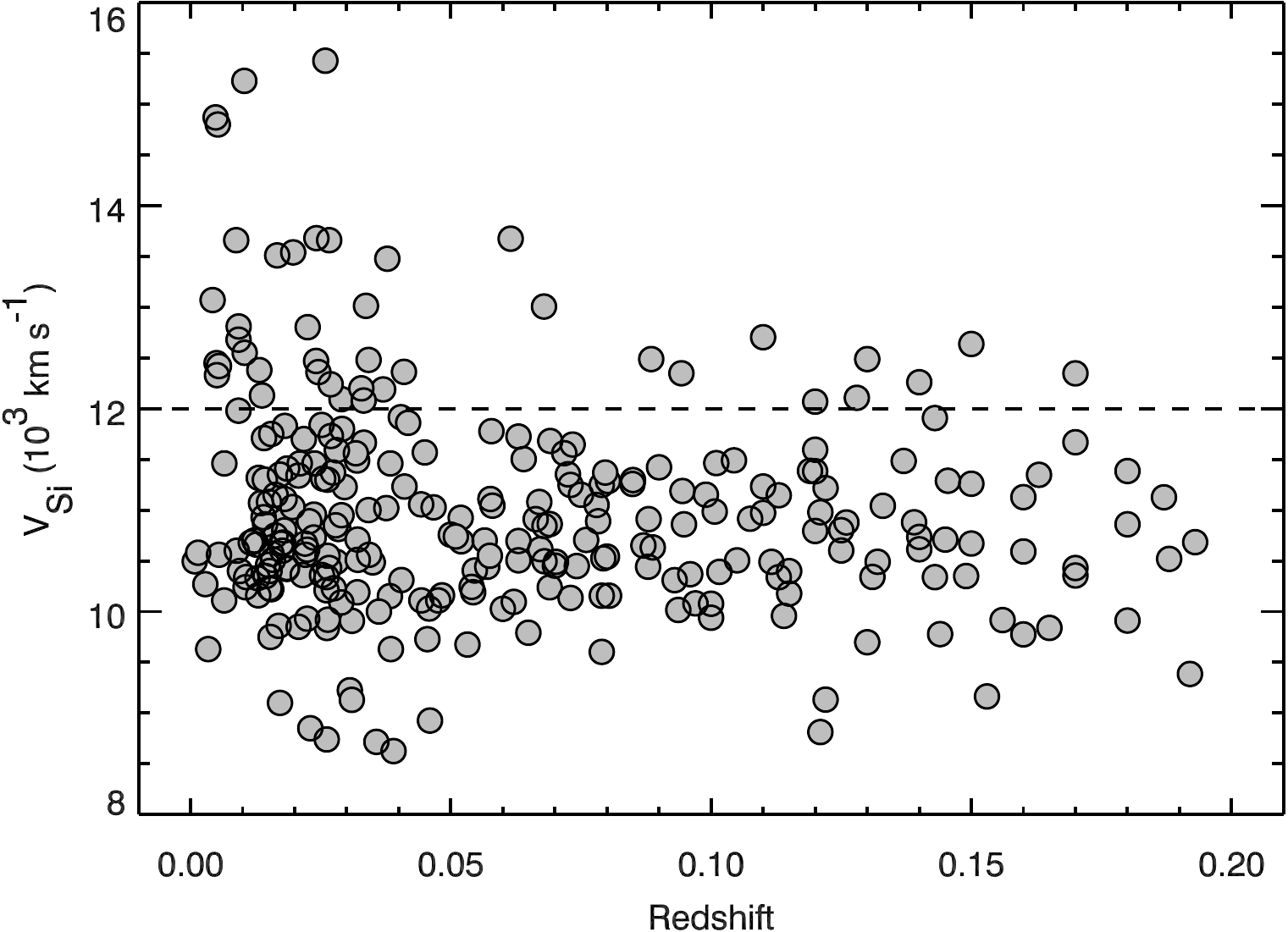}
	\end{tabular}
        \caption{The \Siii\ velocities (\vsiii) as a function of redshift. The dashed line represents the
        the criterion used to split the high-\vsiii\ and normal-\vsiii\ SNe~Ia 
        (i.e., $\vsiii=12000\,\mathrm{km\,s^{-1}}$).}
        \label{z-vsi}
\end{figure}

Our results also have significant implications on cosmology. \citet*{2011ApJ...729...55F} found high-\vsiii\ SNe~Ia are intrinsically redder (in terms of $B-V$ at maximum light) than normal-\vsiii\ SNe~Ia. By accounting for this color difference, they reduced the scatter in Hubble residuals (HRs). Also recently, \citet{2020MNRAS.tmp..548S} determined a \about3-$\sigma$ HR step between high-\vsiii\ and normal-\vsiii\ SNe~Ia, with high-\vsiii\ SNe~Ia having more negative HRs than the normal-\vsiii\ SNe~Ia. These results are indicative that the ejecta velocity can be used to improve SN~Ia distances.

Moreover, if the fraction of high-\vsiii\ to normal-\vsiii\ SNe~Ia changes with redshift, it could introduce significant bias on our cosmological analysis when assuming a single intrinsic color for SN~Ia \citep{2011ApJ...729...55F}. Given the strong preference of metal-rich environments for high-\vsiii\ SNe~Ia, we would expect an evolution on the number of discovered high-\vsiii\ SNe~Ia with redshift, with a decreased rate towards higher redshifts. Fig.~\ref{z-vsi} shows the \vsiii\ as the function of redshift with our sample. We find some evidence that the ratio of high-\vsiii\ to normal-\vsiii\ SNe~Ia tend to be higher at lower redshifts. At $z<0.1$, \about16 percent (13 percent if using untargeted PTF sample only) of SNe in our sample are high-\vsiii\ SNe, while only \about11 percent of the SNe are high-\vsiii\ SNe at $z>0.1$. It is also obvious that the extremely high-\vsiii\ SNe~Ia (e.g., $\vsiii\gtrsim13\,000$\,\kms) are hardly found at $z>0.1$.

However, the selection effect might play a role for SNe discovered at $z>0.1$ in our sample \citep[for a discussion, see][]{2014MNRAS.438.1391P}. For example, we would expect a bias if the high-\vsiii\ SNe~Ia are significantly brighter or fainter than normal-\vsiii\ SNe~Ia. Using $\deltam$($B$) as the proxy of SN~Ia brightness, we determine a mean $\deltam$($B$) of $1.12\pm0.33$\,mag and $1.12\pm0.22$\,mag for normal-\vsiii\ and high-\vsiii\ SNe~Ia, respectively. This indicates (on average) they are not different in brightness. Thus, there is unlikely a Malmquist bias (at least) on the fraction of high-\vsiii\ to normal-\vsiii\ SNe~Ia at higher redshifts. We also note that the high-\vsiii\ SNe~Ia tend to show less dispersion in $\deltam$($B$) than that of normal-\vsiii\ SNe~Ia. It is unclear if this trend is intrinsic to the explosion or simply due to the smaller sample size of high-\vsiii\ SNe~Ia.

Another caveat is the difficulty in finding SNe on very bright galaxy backgrounds, where the contrast of the SN over the host galaxy is low. Since high-\vsiii\ SNe~Ia tend to be found in massive galaxies and are more concentrated in the inner regions of their hosts, they may be more difficult to be fond at higher redshifts. However, this is only an issue with modern image subtraction techniques when the SN brightness drops to $<10\%$ of that of the host background \citep[e.g.,][]{2010AJ....140..518P}. The future analysis with data taken from higher-redshift surveys (e.g., Pan-STARRS1; Pan et al., in preparation) will be necessary to constrain the potential evolution of high-\vsiii\ SNe~Ia.

\section{Conclusions}
\label{sec:conclusions}
In this work, we investigate the relation between photospheric \Siii\ velocities (\vsiii) and host-galaxy properties of SN~Ia. A more stringent criterion on the phase of SN spectra is adopted to distinguish between normal-\vsiii\ and high-\vsiii\ SNe~Ia. We find the high-\vsiii\ SNe~Ia are likely formed from a distinct population which favors massive host environments. This is further supported by the direct measurements on host gas-phase metallicities. Although opacity may have some effect (due to progenitor metallicity), we argue the difference in photospheric velocities between high-\vsiii\ and normal-\vsiii\ SNe~Ia is mainly caused by different explosion mechanisms. 

Theoretical studies suggested the high-\vsiii\ SNe~Ia may originate from sub-Chandrasekhar explosions. This is consistent with our results. At a given mass, stars of higher metallicities generally produce less massive WDs. This may increase the chance for them to explode under sub-Chandrasekhar mass. Nevertheless, the detailed investigation is still needed to evaluate the effect of progenitor metallicity on such models.

Previous studies also suggested the high-\vsiii\ and normal-\vsiii\ SNe~Ia could be formed via single degenerate and double degenerate scenarios, respectively. However, we find high-\vsiii\ SNe~Ia do not tend to originate from younger populations than that of normal-\vsiii\ SNe~Ia. We argue the metallicity is the only important factor in forming high-\vsiii\ SNe~Ia. 

Our results also imply potential evolution of high-\vsiii\ SNe~Ia. We would expect less high-\vsiii\ SNe~Ia to be discovered at higher redshifts when the Universe is more metal-poor than present. This evolution could introduce a bias on our cosmological analysis given that high-\vsiii\ and normal-\vsiii\ SNe~Ia tend to have intrinsically different colors. Future spectroscopic studies using higher-redshift dataset will be critical to measure this evolution effect.

\begin{table*}
\centering
\caption{Summary of our sample in this work.}
\begin{tabular}{lcccc}
\hline\hline
SN Name         & Redshift & \vsiii\ & log\,$M_{\rm stellar} $ & $\rm12+log\,(O/H)$\\
                &              & (\kms) & (M$_{\odot}$)           &                \\
\hline
PTF09bai & 0.180 & $11385\pm 18$ & $10.518^{+0.104}_{-0.279}$ & \nodata \\
PTF09bj & 0.144 & $ 9777\pm 50$ & $10.495^{+0.324}_{-0.123}$ & \nodata \\
PTF09djc & 0.034 & $13013\pm 16$ & $10.771^{+0.008}_{-0.176}$ & \nodata \\
PTF09dlc & 0.067 & $10615\pm  9$ & $ 8.923^{+0.138}_{-0.433}$ & $7.874\pm0.023$ \\
PTF09dnl & 0.024 & $10955\pm  4$ & $ 8.056^{+0.062}_{-0.053}$ & \nodata \\
PTF09dnp & 0.037 & $12189\pm  7$ & $10.586^{+0.143}_{-0.033}$ & $9.121\pm0.013$ \\
PTF09dqt & 0.113 & $11146\pm 22$ & $ 9.357^{+0.055}_{-0.011}$ & \nodata \\
PTF09dxo & 0.052 & $10924\pm 10$ & $10.364^{+0.449}_{-0.157}$ & $8.865\pm0.014$ \\
PTF09e & 0.149 & $10349\pm 38$ & $ 9.312^{+0.154}_{-0.939}$ & \nodata \\
PTF09fox & 0.072 & $11560\pm 11$ & $10.309^{+0.033}_{-0.034}$ & $8.706\pm0.005$ \\
PTF09foz & 0.054 & $10190\pm  8$ & $10.540^{+0.039}_{-0.147}$ & \nodata \\
PTF09gn & 0.139 & $10880\pm 72$ & $ 9.558^{+0.124}_{-0.215}$ & \nodata \\
PTF09gul & 0.072 & $11352\pm 33$ & $10.988^{+0.145}_{-0.011}$ \nodata \\
PTF09h & 0.121 & $10979\pm 31$ & $10.859^{+0.074}_{-0.140}$ & \nodata \\
PTF09ib & 0.122 & $11216\pm 15$ & $ 9.854^{+0.177}_{-0.303}$ & \nodata \\
PTF09isn & 0.101 & $11463\pm 37$ & $ 9.315^{+0.128}_{-0.015}$ & \nodata \\
PTF09s & 0.046 & $ 8926\pm 14$ & $ 9.308^{+0.104}_{-0.355}$ & \nodata \\
PTF09v & 0.119 & $11386\pm 37$ & $ 8.526^{+0.154}_{-0.173}$ & \nodata \\
PTF10aaea & 0.160 & $ 9775\pm147$ & $10.923^{+0.232}_{-0.125}$ & \nodata \\
PTF10abjv & 0.076 & $10705\pm 27$ & $10.024^{+0.039}_{-0.079}$ & \nodata \\
PTF10acqp & 0.170 & $12347\pm132$ & $10.980^{+0.167}_{-0.085}$ & \nodata \\
PTF10bhw & 0.110 & $11233\pm 25$ & $11.423^{+0.049}_{-0.239}$ & \nodata \\
PTF10cmj & 0.112 & $10490\pm 30$ & $11.117^{+0.064}_{-0.057}$ & \nodata \\
PTF10cwm & 0.079 & $10525\pm 18$ & $10.575^{+0.072}_{-0.194}$ & $8.693\pm0.141$ \\
PTF10cxk & 0.018 & $10938\pm 13$ & $ 9.147^{+0.090}_{-0.083}$ & \nodata \\
PTF10duy & 0.079 & $ 9602\pm 15$ & $ 9.242^{+0.074}_{-0.047}$ & \nodata \\
PTF10duz & 0.064 & $11502\pm 33$ & $10.206^{+0.111}_{-0.044}$ & $8.459\pm0.010$ \\
PTF10fej & 0.110 & $12703\pm 82$ & $10.886^{+0.270}_{-0.096}$ & $8.897\pm0.054$ \\
PTF10fj & 0.050 & $10757\pm 19$ & $10.906^{+0.137}_{-0.009}$ & \nodata \\
PTF10fxe & 0.099 & $11152\pm155$ & $ 9.391^{+0.206}_{-0.149}$ & \nodata \\
PTF10fxl & 0.030 & $11227\pm 12$ & $10.740^{+0.013}_{-0.165}$ & $8.785\pm0.003$ \\
PTF10fxp & 0.104 & $11490\pm 17$ & $11.668^{+0.039}_{-0.117}$ & \nodata \\
PTF10fxq & 0.107 & $10914\pm 34$ & $10.971^{+0.079}_{-0.249}$ & $8.923\pm0.123$ \\
PTF10fyl & 0.055 & $10405\pm  9$ & $10.312^{+0.004}_{-0.180}$ & \nodata \\
PTF10glo & 0.075 & $11151\pm 31$ & $ 9.226^{+0.083}_{-0.029}$ & \nodata \\
PTF10gnj & 0.078 & $10889\pm 19$ & $10.273^{+0.098}_{-0.120}$ & $8.586\pm0.107$ \\
PTF10goo & 0.087 & $10651\pm 31$ & $10.763^{+0.106}_{-0.210}$ & $8.690\pm0.114$ \\
PTF10gop & 0.097 & $10080\pm 33$ & $ 8.307^{+0.753}_{-1.440}$ & \nodata \\
PTF10goq & 0.088 & $10910\pm 59$ & $10.557^{+0.086}_{-0.215}$ & \nodata \\
PTF10hdm & 0.165 & $ 9839\pm230$ & $ 9.743^{+0.106}_{-0.030}$ & \nodata \\
PTF10hld & 0.038 & $13477\pm 24$ & $ 9.964^{+0.037}_{-0.090}$ & \nodata \\
PTF10jab & 0.187 & $11126\pm 16$ & $11.346^{+0.132}_{-0.014}$ & \nodata \\
PTF10lot & 0.022 & $12802\pm 16$ & $ 9.902^{+0.079}_{-0.061}$ & \nodata \\
PTF10mtd & 0.079 & $11248\pm 15$ & $ 9.875^{+0.101}_{-0.065}$ & \nodata \\
PTF10mwb & 0.031 & $ 9909\pm  4$ & $ 9.276^{+0.208}_{-0.286}$ & $7.954\pm0.041$ \\
PTF10ncy & 0.130 & $ 9697\pm 45$ & $10.003^{+0.226}_{-0.031}$ & \nodata \\
PTF10ncz & 0.170 & $10429\pm 95$ & $ 9.364^{+0.266}_{-0.025}$ & \nodata \\
PTF10nda & 0.101 & $10985\pm 26$ & $11.374^{+0.049}_{-0.216}$ & \nodata \\
PTF10nhu & 0.153 & $ 9163\pm132$ & $10.212^{+0.252}_{-0.131}$ & \nodata \\
PTF10nnh & 0.150 & $10670\pm106$ & $10.856^{+0.261}_{-0.065}$ & \nodata \\
PTF10nvh & 0.068 & $10843\pm  8$ & $ 9.950^{+0.071}_{-0.117}$ & \nodata \\
PTF10oth & 0.145 & $11289\pm 35$ & $11.237^{+0.174}_{-0.103}$ & \nodata \\
PTF10pvh & 0.105 & $10504\pm 39$ & $ 9.833^{+0.179}_{-0.247}$ & \nodata \\
PTF10pvi & 0.080 & $11286\pm 46$ & $10.326^{+0.079}_{-0.178}$ & $8.668\pm0.066$ \\
PTF10qhp & 0.032 & $11488\pm 25$ & $10.889^{+0.025}_{-0.165}$ & $8.448\pm0.080$ \\
PTF10qjl & 0.058 & $11109\pm  5$ & $ 8.671^{+0.020}_{-0.299}$ & $7.926\pm0.052$ \\
PTF10qjq & 0.028 & $10813\pm 12$ & $10.059^{+0.087}_{-0.044}$ & $8.711\pm0.003$ \\
\hline
\label{sample1}
\end{tabular}
\end{table*}

\begin{table*}
\centering
\caption{Summary of our sample in this work (continued).}
\begin{tabular}{lcccc}
\hline\hline
SN Name         & Redshift & \vsiii\ & log\,$M_{\rm stellar} $ & $\rm12+log\,(O/H)$\\
                &              & (\kms) & (M$_{\odot}$)           &                \\
\hline
PTF10qkf & 0.080 & $10156\pm 55$ & $10.445^{+0.115}_{-0.161}$ & $8.428\pm0.026$ \\
PTF10qky & 0.074 & $10448\pm 13$ & $10.559^{+0.094}_{-0.116}$ & $8.663\pm0.005$ \\
PTF10qny & 0.033 & $11664\pm 12$ & $10.340^{+0.484}_{-0.160}$ & $8.547\pm0.014$ \\
PTF10qsc & 0.088 & $10441\pm 31$ & $ 9.762^{+0.137}_{-0.378}$ & $7.854\pm0.082$ \\
PTF10qwg & 0.068 & $13006\pm 16$ & $10.089^{+0.151}_{-0.157}$ & $8.749\pm0.006$ \\
PTF10qyq & 0.160 & $10592\pm 99$ & $ 9.445^{+0.212}_{-0.050}$ & \nodata \\
PTF10rab & 0.085 & $11296\pm 70$ & $ 7.920^{+1.043}_{-0.114}$ & $7.811\pm0.005$ \\
PTF10ran & 0.160 & $11127\pm 57$ & $11.185^{+0.088}_{-0.147}$ & \nodata \\
PTF10rhi & 0.085 & $11257\pm 35$ & $10.014^{+0.077}_{-0.364}$ & \nodata \\
PTF10tce & 0.041 & $12362\pm 12$ & $10.542^{+0.161}_{-0.155}$ & $8.745\pm0.053$ \\
PTF10tqy & 0.045 & $11568\pm 11$ & $10.101^{+0.162}_{-0.005}$ & \nodata \\
PTF10trs & 0.073 & $11248\pm  2$ & $10.824^{+0.062}_{-0.070}$ & \nodata \\
PTF10trw & 0.170 & $10355\pm 40$ & $ 9.387^{+0.259}_{-0.024}$ & \nodata \\
PTF10twd & 0.073 & $11646\pm 23$ & $ 9.884^{+0.089}_{-0.055}$ & $8.365\pm0.006$ \\
PTF10ucl & 0.080 & $10541\pm 15$ & $10.683^{+0.561}_{-0.020}$ & \nodata \\
PTF10ufj & 0.073 & $10134\pm 10$ & $10.789^{+0.428}_{-0.051}$ & \nodata \\
PTF10urn & 0.110 & $10972\pm 76$ & $ 9.543^{+0.262}_{-0.487}$ & \nodata \\
PTF10vfo & 0.088 & $12488\pm 63$ & $10.751^{+0.011}_{-0.022}$ & \nodata \\
PTF10viq & 0.034 & $12480\pm 10$ & $10.792^{+0.201}_{-0.417}$ & $8.900\pm0.013$ \\
PTF10wnm & 0.066 & $10910\pm 11$ & $10.548^{+0.045}_{-0.116}$ & $8.749\pm0.044$ \\
PTF10wnq & 0.069 & $11681\pm 56$ & $11.341^{+0.107}_{-0.543}$ & \nodata \\
PTF10wov & 0.096 & $10368\pm 27$ & $10.165^{+0.276}_{-0.076}$ & \nodata \\
PTF10wri & 0.120 & $11595\pm 20$ & $11.777^{+0.101}_{-0.204}$ & \nodata \\
PTF10xeb & 0.122 & $ 9132\pm 62$ & $ 7.316^{+0.665}_{-0.091}$ & \nodata \\
PTF10xir & 0.052 & $10692\pm 19$ & $11.025^{+0.428}_{-0.178}$ & \nodata \\
PTF10xtp & 0.102 & $10391\pm103$ & $10.419^{+0.099}_{-0.133}$ & \nodata \\
PTF10ygr & 0.115 & $10177\pm114$ & $ 9.698^{+0.200}_{-0.022}$ & \nodata \\
PTF10ygu & 0.026 & $15428\pm 11$ & $10.911^{+0.491}_{-0.161}$ & $8.672\pm0.003$ \\
PTF10yux & 0.058 & $11777\pm 19$ & $10.764^{+0.006}_{-0.066}$ & \nodata \\
PTF10zai & 0.036 & $ 8715\pm 59$ & $10.822^{+0.563}_{-0.067}$ & $9.005\pm0.203$ \\
PTF10zak & 0.040 & $11915\pm 10$ & $10.757^{+0.400}_{-0.157}$ & $8.997\pm0.034$ \\
PTF10zbn & 0.114 & $ 9957\pm249$ & $11.189^{+0.094}_{-0.160}$ & \nodata \\
PTF10zgy & 0.044 & $11056\pm 40$ & $11.238^{+0.000}_{-0.000}$ & \nodata \\
PTF11apk & 0.041 & $10310\pm 10$ & $10.914^{+0.007}_{-0.164}$ & \nodata \\
PTF11bjk & 0.140 & $10613\pm 61$ & $10.954^{+0.064}_{-0.203}$ & \nodata \\
PTF11blu & 0.068 & $10495\pm 35$ & $ 9.963^{+0.016}_{-0.081}$ & $8.605\pm0.165$ \\
PTF11byi & 0.039 & $ 8625\pm 25$ & $10.527^{+0.077}_{-0.581}$ & $8.808\pm0.094$ \\
PTF11ctn & 0.079 & $10156\pm 11$ & $ 9.312^{+0.313}_{-0.100}$ & \nodata \\
PTF11cyv & 0.115 & $10398\pm 46$ & $ 9.851^{+0.125}_{-0.095}$ & \nodata \\
PTF11deg & 0.063 & $11724\pm 31$ & $ 9.452^{+0.097}_{-0.074}$ & \nodata \\
PTF11dws & 0.150 & $12637\pm 45$ & $11.174^{+0.095}_{-0.308}$ & \nodata \\
PTF11dzm & 0.041 & $11232\pm 59$ & $10.090^{+0.007}_{-0.163}$ & \nodata \\
PTF11eot & 0.090 & $11421\pm 50$ & $ 8.600^{+0.105}_{-0.135}$ & \nodata \\
PTF11fjw & 0.193 & $10684\pm 66$ & $10.892^{+0.216}_{-0.136}$ & \nodata \\
PTF11for & 0.128 & $12106\pm 83$ & $ 9.638^{+0.033}_{-0.083}$ & \nodata \\
PTF11gdh & 0.026 & $ 9838\pm 13$ & $10.390^{+0.190}_{-0.099}$ & $8.743\pm0.025$ \\
PTF11gin & 0.163 & $11347\pm 42$ & $ 9.626^{+0.255}_{-0.546}$ & \nodata \\
PTF11gjb & 0.125 & $10601\pm 42$ & $10.926^{+0.052}_{-0.131}$ & \nodata \\
PTF11glq & 0.156 & $ 9916\pm166$ & $11.360^{+0.088}_{-0.019}$ & \nodata \\
PTF11gnj & 0.131 & $10341\pm 61$ & $10.387^{+0.114}_{-0.032}$ & \nodata \\
PTF11khk & 0.031 & $ 9224\pm 11$ & $10.704^{+0.058}_{-0.120}$ & $8.926\pm0.034$ \\
PTF11kod & 0.188 & $10519\pm297$ & $ 8.741^{+0.912}_{-0.220}$ & \nodata \\
PTF11kpb & 0.143 & $10339\pm 19$ & $11.675^{+0.091}_{-0.014}$ & \nodata \\
PTF11kqm & 0.126 & $10878\pm 21$ & $10.428^{+0.241}_{-0.030}$ & \nodata \\
PTF11kx & 0.047 & $11027\pm 40$ & $10.315^{+0.169}_{-0.152}$ & $8.505\pm0.007$ \\
PTF11lbc & 0.150 & $11260\pm 44$ & $ 7.967^{+1.320}_{-0.848}$ & \nodata \\
PTF11lmz & 0.061 & $13674\pm 21$ & $11.206^{+0.074}_{-0.203}$ & \nodata \\
\hline
\label{sample2}
\end{tabular}
\end{table*}

\begin{table*}
\centering
\caption{Summary of our sample in this work (continued).}
\begin{tabular}{lcccc}
\hline\hline
SN Name         & Redshift & \vsiii\ & log\,$M_{\rm stellar} $ & $\rm12+log\,(O/H)$\\
                &              & (\kms) & (M$_{\odot}$)           &                \\
\hline
PTF11opu & 0.065 & $ 9791\pm 47$ & $ 9.797^{+0.127}_{-0.094}$ & $8.076\pm0.050$ \\
PTF11qpc & 0.089 & $10632\pm 23$ & $10.312^{+0.217}_{-0.239}$ & $8.643\pm0.123$ \\
PTF11qvh & 0.133 & $11042\pm 12$ & $ 9.800^{+0.171}_{-0.013}$ & \nodata \\
PTF11rke & 0.094 & $12348\pm 35$ & $ 9.968^{+0.066}_{-0.136}$ & \nodata \\
PTF11rpc & 0.143 & $11907\pm 49$ & $11.315^{+0.225}_{-0.183}$ & \nodata \\
PTF11yp & 0.121 & $ 8812\pm 41$ & $11.447^{+0.068}_{-0.313}$ & \nodata \\
PTF12awi & 0.045 & $ 9727\pm 48$ & $10.503^{+0.133}_{-0.108}$ & $8.764\pm0.052$ \\
PTF12cdb & 0.120 & $12068\pm 20$ & $10.877^{+0.212}_{-0.056}$ & \nodata \\
PTF12cjg & 0.067 & $11080\pm 17$ & $ 9.916^{+0.105}_{-0.131}$ & $8.615\pm0.142$ \\
PTF12cyd & 0.170 & $11668\pm 12$ & $ 9.118^{+0.446}_{-0.073}$ & \nodata \\
PTF12czu & 0.145 & $10709\pm151$ & $ 9.005^{+0.021}_{-0.138}$ & \nodata \\
PTF12dgy & 0.180 & $ 9911\pm 23$ & $10.919^{+0.367}_{-0.175}$ & \nodata \\
PTF12dhb & 0.057 & $10701\pm 12$ & $10.373^{+0.030}_{-0.072}$ & $8.618\pm0.032$ \\
PTF12dhl & 0.057 & $10441\pm 38$ & $11.001^{+0.111}_{-0.012}$ & \nodata \\
PTF12dhv & 0.140 & $12260\pm 80$ & $10.442^{+0.144}_{-0.160}$ & \nodata \\
PTF12dst & 0.192 & $ 9385\pm 24$ & $ 9.939^{+0.079}_{-0.019}$ & \nodata \\
PTF12dwm & 0.053 & $ 9675\pm 10$ & $10.538^{+0.159}_{-0.200}$ & $8.765\pm0.074$ \\
PTF12dxm & 0.063 & $10505\pm  9$ & $11.033^{+0.005}_{-0.005}$ & \nodata \\
PTF12egr & 0.132 & $10490\pm 46$ & $11.520^{+0.126}_{-0.013}$ & \nodata \\
PTF12fhn & 0.125 & $10801\pm 76$ & $11.281^{+0.142}_{-0.011}$ & \nodata \\
PTF12fsd & 0.069 & $10862\pm 25$ & $10.107^{+0.098}_{-0.117}$ & $8.398\pm0.063$ \\
PTF12giy & 0.029 & $12093\pm177$ & $10.835^{+0.177}_{-0.008}$ & $8.826\pm0.096$ \\
PTF12gnw & 0.100 & $10077\pm 14$ & $ 8.816^{+0.187}_{-0.234}$ & \nodata \\
SN~1989M & 0.005 & $12330\pm 50$ & $10.539^{+0.627}_{-0.144}$ & \nodata \\
SN~1994S & 0.015 & $10400\pm 50$ & $11.113^{+0.036}_{-0.520}$ & $8.943\pm0.013$ \\
SN~1997Y & 0.016 & $10510\pm 50$ & $11.018^{+0.067}_{-0.045}$ & $9.023\pm0.012$ \\
SN~1998dk & 0.013 & $12380\pm 50$ & $ 9.982^{+0.033}_{-0.186}$ & \nodata \\
SN~1998es & 0.011 & $10240\pm 50$ & $10.299^{+0.052}_{-0.010}$ & \nodata \\
SN~1999aa & 0.014 & $10350\pm 50$ & $10.533^{+0.132}_{-0.013}$ & $9.035\pm0.027$ \\
SN~1999ac & 0.009 & $10400\pm 50$ & $10.254^{+0.504}_{-0.128}$ & \nodata \\
SN~1999da & 0.013 & $10660\pm 50$ & $10.886^{+0.039}_{-0.120}$ & \nodata \\
SN~1999dq & 0.014 & $10860\pm 50$ & $10.941^{+0.422}_{-0.118}$ & \nodata \\
SN~1999gd & 0.018 & $10420\pm 50$ & $10.361^{+0.168}_{-0.044}$ & \nodata \\
SN~2000cp & 0.034 & $11000\pm 50$ & $ 9.976^{+0.459}_{-0.071}$ & $8.841\pm0.026$ \\
SN~2000dn & 0.032 & $10190\pm 50$ & $11.081^{+0.010}_{-0.177}$ & \nodata \\
SN~2001bp & 0.095 & $10860\pm 50$ & $10.602^{+0.065}_{-0.165}$ & $8.795\pm0.032$ \\
SN~2001da & 0.017 & $11350\pm 50$ & $10.633^{+0.440}_{-0.185}$ & \nodata \\
SN~2001ep & 0.013 & $10160\pm 50$ & $10.215^{+0.023}_{-0.117}$ & \nodata \\
SN~2001fe & 0.014 & $11070\pm 50$ & $10.341^{+0.109}_{-0.077}$ & \nodata \\
SN~2002aw & 0.026 & $10210\pm 50$ & $10.931^{+0.051}_{-0.076}$ & $8.976\pm0.029$ \\
SN~2002bf & 0.024 & $13680\pm 50$ & $10.672^{+0.004}_{-0.004}$ & \nodata \\
SN~2002bo & 0.004 & $13070\pm 50$ & $10.779^{+0.054}_{-0.449}$ & \nodata \\
SN~2002cd & 0.010 & $15230\pm 50$ & $10.971^{+0.124}_{-0.010}$ & \nodata \\
SN~2002dk & 0.019 & $10430\pm 50$ & $11.656^{+0.048}_{-0.151}$ & \nodata \\
SN~2002eb & 0.028 & $10230\pm 50$ & $10.976^{+0.164}_{-0.423}$ & \nodata \\
SN~2002eu & 0.038 & $11020\pm 50$ & $10.648^{+0.493}_{-0.070}$ & \nodata \\
SN~2002ha & 0.014 & $10930\pm 50$ & $11.075^{+0.481}_{-0.198}$ & \nodata \\
SN~2003U & 0.026 & $11300\pm 50$ & $ 9.670^{+0.614}_{-0.377}$ & $8.828\pm0.021$ \\
SN~2003Y & 0.017 & $ 9860\pm 50$ & $10.728^{+0.153}_{-0.037}$ & \nodata \\
SN~2003cq & 0.033 & $12080\pm 50$ & $11.884^{+0.064}_{-0.111}$ & \nodata \\
SN~2003he & 0.025 & $11310\pm 50$ & $10.335^{+0.545}_{-0.069}$ & \nodata \\
SN~2004dt & 0.020 & $13540\pm 50$ & $10.875^{+0.020}_{-0.144}$ & \nodata \\
SN~2004gs & 0.027 & $10430\pm 50$ & $10.740^{+0.137}_{-0.032}$ & \nodata \\
SN~2005M & 0.022 & $10670\pm 60$ & $ 9.898^{+0.081}_{-0.030}$ & $8.504\pm0.006$ \\
SN~2005W & 0.009 & $10600\pm 50$ & $10.473^{+0.007}_{-0.154}$ & \nodata \\
SN~2005ao & 0.038 & $11460\pm 50$ & $10.915^{+0.161}_{-0.033}$ & \nodata \\
SN~2005ag & 0.079 & $11370\pm 50$ & \nodata & $9.250\pm0.021$ \\
SN~2005bc & 0.012 & $10700\pm 50$ & $10.383^{+0.018}_{-0.120}$ & $9.054\pm0.007$ \\
\hline
\label{sample3}
\end{tabular}
\end{table*}

\begin{table*}
\centering
\caption{Summary of our sample in this work (continued).}
\begin{tabular}{lcccc}
\hline\hline
SN Name         & Redshift & \vsiii\ & log\,$M_{\rm stellar} $ & $\rm12+log\,(O/H)$\\
                &              & (\kms) & (M$_{\odot}$)           &                \\
\hline
SN~2005cg & 0.032 & $11560\pm 50$ & $ 8.574^{+0.057}_{-0.025}$ & $7.947\pm0.039$ \\
SN~2005er & 0.026 & $ 8740\pm 50$ & $11.478^{+0.492}_{-0.114}$ & \nodata \\
SN~2005eq & 0.029 & $10090\pm 50$ & $11.424^{+0.029}_{-0.452}$ & \nodata \\
SN~2005hj & 0.058 & $10550\pm 50$ & $10.046^{+0.078}_{-0.038}$ & $8.519\pm0.038$ \\
SN~2005ki & 0.020 & $11030\pm 50$ & $11.146^{+0.035}_{-0.120}$ & \nodata \\
SN~2005ms & 0.025 & $11840\pm 50$ & $10.684^{+0.210}_{-0.054}$ & \nodata \\
SN~2006N & 0.014 & $11300\pm 50$ & $10.545^{+0.494}_{-0.143}$ & \nodata \\
SN~2006S & 0.032 & $10710\pm 50$ & $10.628^{+0.006}_{-0.184}$ & \nodata \\
SN~2006bt & 0.032 & $10510\pm 50$ & $11.124^{+0.027}_{-0.155}$ & \nodata \\
SN~2006bz & 0.028 & $10850\pm 50$ & $10.471^{+0.006}_{-0.020}$ & \nodata \\
SN~2006cm & 0.016 & $11150\pm 50$ & $10.913^{+0.069}_{-0.047}$ & \nodata \\
SN~2006cq & 0.048 & $10160\pm 50$ & $11.212^{+0.011}_{-0.141}$ & \nodata \\
SN~2006cs & 0.024 & $10730\pm 50$ & $10.913^{+0.490}_{-0.024}$ & \nodata \\
SN~2006or & 0.021 & $11340\pm 50$ & $11.114^{+0.034}_{-0.117}$ & \nodata \\
SN~2006sr & 0.024 & $12470\pm 50$ & $10.778^{+0.325}_{-0.163}$ & \nodata \\
SN~2007A & 0.018 & $10600\pm 50$ & $10.775^{+0.252}_{-0.241}$ & \nodata \\
SN~2007N & 0.013 & $10330\pm 50$ & $10.376^{+0.017}_{-0.141}$ & \nodata \\
SN~2007O & 0.036 & $10000\pm 50$ & $11.033^{+0.007}_{-0.117}$ & $9.009\pm0.011$ \\
SN~2007af & 0.005 & $10560\pm 50$ & $ 9.562^{+0.181}_{-0.096}$ & \nodata \\
SN~2007ba & 0.038 & $ 9630\pm 50$ & $11.118^{+0.105}_{-0.068}$ & \nodata \\
SN~2007bc & 0.021 & $ 9850\pm 50$ & $10.922^{+0.020}_{-0.000}$ & \nodata \\
SN~2007bz & 0.022 & $11700\pm 50$ & $ 9.504^{+0.085}_{-0.100}$ & $8.329\pm0.012$ \\
SN~2007ci & 0.018 & $11830\pm 50$ & $10.969^{+0.007}_{-0.150}$ & \nodata \\
SN~2007fr & 0.051 & $10740\pm 50$ & $11.770^{+0.032}_{-0.102}$ & \nodata \\
SN~2007gi & 0.005 & $14870\pm 50$ & $10.643^{+0.479}_{-0.151}$ & \nodata \\
SN~2007hj & 0.014 & $11710\pm 50$ & $10.580^{+0.484}_{-0.143}$ & \nodata \\
SN~2007s1 & 0.027 & $11370\pm 50$ & $ 9.998^{+0.143}_{-0.030}$ & \nodata \\
SN~2008Z & 0.021 & $11460\pm 50$ & $ 9.488^{+0.203}_{-0.064}$ & $8.359\pm0.060$ \\
SN~2008ar & 0.026 & $10340\pm 50$ & $10.880^{+0.071}_{-0.512}$ & $8.972\pm0.016$ \\
SN~2008s1 & 0.022 & $10560\pm 50$ & $10.559^{+0.129}_{-0.040}$ & \nodata \\
SN~2008dx & 0.023 & $ 8850\pm 50$ & $10.678^{+0.027}_{-0.136}$ & \nodata \\
SN~2008ec & 0.016 & $10750\pm 50$ & $10.930^{+0.124}_{-0.075}$ & \nodata \\
SN~2009an & 0.009 & $12680\pm 50$ & $10.981^{+0.084}_{-0.000}$ & $9.086\pm0.014$ \\
SN~2009fx & 0.048 & $10110\pm 50$ & $10.000^{+0.020}_{-0.170}$ & \nodata \\
SN~2009ig & 0.009 & $13660\pm 50$ & $10.454^{+0.000}_{-0.175}$ & \nodata \\
SN~2009no & 0.046 & $10030\pm 50$ & $10.319^{+0.041}_{-0.187}$ & $8.776\pm0.028$ \\
SN~2010ex & 0.023 & $10890\pm 50$ & $10.089^{+0.142}_{-0.030}$ & \nodata \\
SN~2010ii & 0.027 & $12240\pm 50$ & $11.142^{+0.013}_{-0.144}$ & \nodata \\
SN~2010iw & 0.022 & $10360\pm 50$ & $ 9.731^{+0.066}_{-0.092}$ & \nodata \\
SN~2011ao & 0.011 & $10340\pm 50$ & $ 9.877^{+0.141}_{-0.032}$ & $8.481\pm0.010$ \\
SN~2011by & 0.003 & $10270\pm 50$ & $ 9.354^{+0.014}_{-0.191}$ & $8.624\pm0.022$ \\
SN~2011hb & 0.029 & $11800\pm 50$ & $11.247^{+0.004}_{-0.051}$ & \nodata \\
SN~2011ia & 0.017 & $10680\pm 50$ & \nodata & $8.865\pm0.010$ \\
SN~2012bh & 0.025 & $10360\pm 50$ & $10.538^{+0.406}_{-0.187}$ & \nodata \\
SN~2012cg & 0.001 & $10580\pm 50$ & $ 8.467^{+0.035}_{-0.160}$ & $8.729\pm0.007$ \\
SN~2012da & 0.018 & $11110\pm 50$ & $ 9.580^{+0.025}_{-0.061}$ & \nodata \\
SN~2013di & 0.024 & $11460\pm 50$ & $11.008^{+0.530}_{-0.060}$ & \nodata \\
\hline
\label{sample4}
\end{tabular}
\end{table*}

\section*{acknowledgments}
Y.-C.P. is supported by the East Asian Core Observatories Association (EACOA) Fellowship.

\bibliographystyle{apj}
\bibliography{host_revisit}

\end{document}